\documentstyle[prd,aps,preprint,epsf]{revtex}

\draft

\begin{document}
%@twocolumnfalse\endcsname

\baselineskip 14pt

\draft

\title{\baselineskip 14pt
\hbox to\hsize{\small Submitted to Physical Review D \hfil 
FERMILAB--Pub--95/406-A}
\hbox to\hsize{\small\hfill CERN-TH/96-112}
%\hbox to\hsize{\small\hfill MPA~999}
%\hbox to\hsize{\small\hfill SFB-375/???}
\hbox to\hsize{\small\hfill E-Print astro-ph/9612222}
\vskip0.5cm
A Fresh Look at Axions and SN~1987A}

\author{Wolfgang Keil and Hans-Thomas Janka}
\address{Max-Planck-Institut f\"ur Astrophysik\\
Karl-Schwarzschild-Str.~1, 85740 Garching, Germany}

\author{David~N.~Schramm, G\"unter~Sigl, and
Michael~S.~Turner}
\address{Department of Astronomy \& Astrophysics\\
Enrico Fermi Institute, The University of Chicago, Chicago, IL~~60637-1433\\
NASA/Fermilab Astrophysics Center\\
Fermi National Accelerator Laboratory, Batavia, IL~~60510-0500}

\author{John Ellis}
\address{Theoretical Physics Division, CERN, CH-1211
Geneva 23, Switzerland}

\maketitle

\begin{abstract}
\baselineskip 14pt
We re-examine the very stringent limits on the axion mass based
on the strength and duration of the neutrino signal from
SN~1987A, in the light of new measurements of the axial-vector
coupling strength of nucleons, possible suppression of axion emission
due to many-body effects, and additional emission processes
involving pions. The suppression of axion emission
due to nucleon spin fluctuations induced by many-body
effects degrades previous limits by a factor of about
2. Emission processes involving thermal pions can strengthen
the limits by a factor of 3--4 within a perturbative treatment
that neglects saturation of nucleon spin fluctuations. Inclusion
of saturation effects, however, tends to make the limits less 
dependent on pion abundances. The resulting axion mass limit 
also depends on the precise couplings of the axion and ranges from
$0.5\times 10^{-3} \,{\rm eV}$ to $6\times 10^{-3} \,{\rm eV}$.
\end{abstract}

\pacs{PACS numbers: 14.80.Mz, 97.60.Bw, 21.65.+f, 95.30.Cq}
%\vskip2pc]

\narrowtext

\section{Introduction}

Peccei-Quinn (PQ) symmetry \cite{PQ} continues to be an attractive
solution to the strong $CP$ problem.  However, theory and
laboratory experiment give little guidance on
the PQ sym\-me\-try-\-breaking scale, $f_a$,
and therefore on the mass of the pseudo-Goldstone boson
associated with PQ symmetry, the axion \cite{WW}:
\begin{equation}
m_a \simeq 0.62 \,{\rm eV}\cdot 10^7\,{\rm GeV} /f_a \ .
\label{maxion}
\end{equation}
Astrophysical
and cosmological arguments have been very powerful in excluding
values of the axion mass, allowing only the mass range
from about $10^{-6}\,{\rm eV}$ to about $10^{-3}\,{\rm eV}$ \cite{physrepts}.

Axions in this mass range would have been produced cosmologically
by a coherent, nonthermal mechanism \cite{misalignment}.
Because of this, axions in this mass
range would be cosmologically
significant, have small velocities, and
behave as cold dark matter (CDM), in spite of their small mass.
Depending upon cosmological
and particle physics parameters, axions could
contribute to the closure mass density today for
$m_a \simeq 10^{-6}\,{\rm eV}$ to $10^{-4}\,{\rm eV}$ \cite{omega_a}.
We recall also that the CDM scenario for structure formation is at
present the most promising \cite{cdm}, with axions and neutralinos
being the leading CDM candidates.
If axions provide the bulk of the dark matter,
they must comprise a significant fraction of the dark halo of our own
galaxy \cite{sikivie}, and a large-scale experiment is underway
to detect halo axions of mass $10^{-6}\,{\rm eV}$ to $10^{-5}\,{\rm eV}$ \cite{llnl}.

The most stringent astrophysical
bound on the axion mass is that derived from observations of
neutrinos from SN~1987A, excluding masses greater than about
$10^{-3}\,{\rm eV}$ \cite{sn87axlimit,burrowsetal}.
It is based upon the detection of 19 neutrinos from SN~1987A by the
Kamiokande II (KII) \cite{kii} and Irvine-Michigan-Brookhaven (IMB)
\cite{imb} water-Cherenkov
detectors.  According to the standard model of Type II supernovae,
namely the core collapse of a massive star, these neutrinos were
emitted during the early cooling phase of the nascent neutron
star associated with the appearance of SN~1987A.
Indeed, the observed
neutrino flux and energy spectrum are consistent with
this picture \cite{burrowsreview}.  The emission of axions would have
hastened the cooling process, leading to fewer events over a shorter
time.  The axion's couplings to ordinary matter are proportional
to the axion mass, so consistency with the detected neutrino
burst leads to an upper limit to the axion mass, which is
estimated to be around $10^{-3}\,{\rm eV}$.

A number of questions have been raised about the mass limit based upon
SN~1987A. They include the possible suppression of axion
emission by many-body effects which are likely to be important
in the deep interior of a neutron star \cite{R1,R2,R3}, enhancement of
axion emission due to the possible presence of kaons and thermal pions
at the centers of neutron stars \cite{mayle}, and
the implications of recent measurements of the strengths of
axial-current couplings to nucleons, which may lead to
a significant suppression of the
axion-neutron coupling relative to estimates based on
the naive quark model (NQM) \cite{R3}.

The purpose of this paper is to address these issues; its
outline is as follows. Sect.~II is devoted to a discussion of
the axion-nucleon couplings and a description of the input microphysics 
of the numerical protoneutron star models. The axion emission rates
are reviewed in Sect.~III, including many-body effects, with particular
attention to the possible damping effects of nucleon spin fluctuations.
Sect.~IV discusses cooling models and the impact of axion emission
on the theoretical predictions for the neutrino bursts detected by KII
and IMB. We finish with a summary of our results in Sect.~V.

\section{Input Physics}
\subsection{Axion Couplings\label{axcoup}}

The most important axion-emission process in a hot, young
neutron star is axion brems\-strah\-lung when two nucleons collide.
The rate for this process
depends upon the axion's coupling to nucleons.
The relevant part of the interaction Lagrangian is
\begin{equation}
{\cal L}_{\rm int} = {g_{ai}\over 2m_i}{\bar \psi}_i \gamma_\mu
\gamma_5 \psi_i \, \partial^\mu a \ ,\label{Lagran}
\end{equation}
where the index $i$ denotes a neutron or proton,
and we can write the axion-nucleon couplings $g_{ai}$ in
the form
\begin{equation}
g_{ai} \equiv C_i m_i/(f_{\rm PQ}/N) \equiv C_i m_i/f_a
\ .
\label{defc}
\end{equation}
$N$ is the color anomaly of the PQ symmetry.
The dimensionless couplings $C_i$ of Eq.~(\ref{defc})
are of order unity, and related by generalized
Goldberger-Treiman relations to nucleon axial-vector
current matrix elements via the PQ
charges $x_q$ of the light quarks $q = u,d,s$:
\begin{eqnarray}
C_p & = & [x_u -1/(1+z+w)]\Delta u + [x_d -z/(1+z+w)]\Delta d
        + [x_s -w/(1+z+w)]\Delta s ;\nonumber \\
C_n & = & [x_u -1/(1+z+w)]\Delta d + [x_d -z/(1+z+w)]\Delta u
        + [x_s -w/(1+z+w)]\Delta s ; \label{coupling}
\end{eqnarray}
where we use for the light-quark mass ratios
\begin{equation}
z \equiv m_u/m_d \simeq 0.565\,,\ \, w \equiv m_u/m_s \simeq 0.029
\ .
\label{zwvals}
\end{equation}
The PQ charges of the quarks are
model dependent: they vanish for the KVSZ axion \cite{KVSZ},
$x_u=x_d=x_s=0$, whilst for the DFSZ axion\cite{DFSZ}
they can be written in terms of an angle
$\beta$ which is related to a ratio of Higgs
vacuum expectation values:
\begin{equation}
x_u=\sin^2\beta /3, \, x_d=x_s =\cos^2\beta /3
\label{DFSZx}
\end{equation}
The $\Delta q$ in Eq.~(\ref{coupling}) quantify the
axial-vector current couplings to the proton:
\begin{equation}
\Delta q \,S_\mu \equiv \langle p|{\bar q} \gamma_\mu\gamma_5 q|p\rangle
\end{equation}
where $S_\mu$ is the proton spin. Similar expressions hold for
the neutron, with matrix elements related by an isospin
reflection: $\Delta u_n = \Delta d, \Delta d_n = \Delta u$.

The $\Delta q$ are non-perturbative quantities whose values must
be taken from experiment. Neutron $\beta$ decay and
isospin invariance  constrain
$\Delta u - \Delta d \equiv g_A \simeq 1.25$, whilst
hyperon $\beta$ decays and flavor $SU(3)$ symmetry for
the baryon octet yield $\Delta u +\Delta d - 2\Delta s \simeq 0.682$.
The best determinations of the third combination of the $\Delta q$
are obtained from spin-dependent deep-inelastic electron
and muon scattering off nucleons.  Recent analyses
give \cite{deltaq}:
\begin{eqnarray}
\Delta u & = & \phantom{-} 0.80 \pm 0.04 \pm 0.04\,;\nonumber \\
\Delta d & = & -0.46 \pm 0.04 \pm 0.04\,;\label{notNQM} \\
\Delta s & = & -0.12 \pm 0.04 \pm 0.04\,; \nonumber
\end{eqnarray}
where the first error is a statistical error,
and the second is an estimated systematic error.
The values in Eq.~(\ref{notNQM}) are quite different from those
estimated in the NQM, which does not estimate axial-current
matrix elements reliably.
As can be seen in Eq.~(\ref{coupling}), the
measurements of Eq.~(\ref{notNQM}) and the PQ charges of the light 
quarks determine the axion-nucleon couplings. Values
for the KSVZ and DFSZ axions are shown in Fig.~\ref{F1}.

It is perhaps worth warning the reader at this point that
the above values of the $\Delta q$ have been obtained for
nucleons that are free, or in light nuclei. It is possible that
the axial-vector current couplings may be different for
nucleons in a dense medium. However, large
effects on these couplings are not seen in
conventional nuclear physics, and we believe that other
many-body effects on the axion emission rate are dominant,
as we discuss in Sect.~\ref{manyb}.

\subsection{Protoneutron Star Physics\label{pns-ph}}

In order to obtain limits on the axion production in 
SN~1987A and thus on the axion mass,
we performed numerical simulations of the evolution
of newly formed neutron stars. The simulations started a few
milliseconds after core bounce and supernova shock formation and
followed the Kelvin-Helmholtz cooling by neutrino emission
until the lepton-rich and hot protoneutron star had evolved 
to the final cold and neutronized state several ten seconds
later. Cooling sequences 
were computed with and without axion emission included, and
the corresponding neutrino luminosities and spectra were used to
derive predictions for the associated KII and IMB
detector signals.

The initial models of the protoneutron star were constructed
with profiles of the electron concentration $Y_e = n_e/n_b$ 
($n_e$ is the electron number density, $n_b$ the baryon number
density) and of the
temperature $T$ that were very close to those obtained in detailed 
hydrodynamical calculations of stellar core collapse 
\cite{wilson}. The collapsed stellar core right after core bounce 
is gravitationally only weakly bound. This is expressed by the fact
that the ratio of the gravitational mass at the beginning of the 
simulation, $M_{\rm ns,g}^{\rm i}$,
to the baryonic mass, $M_{\rm ns,b}$, is only slightly less than
unity. We scaled the temperature profiles from given numerical 
models of post-collapse cores such that 
$M_{\rm ns,g}^{\rm i}/M_{\rm ns,b} = 0.97$ in all of our initial setups.
The baryonic mass of the neutron star depends on the mass of 
the progenitor star of the supernova. Simulations
with varied protoneutron star mass \cite{keil,keiletal}
showed that the neutrino signal from SN~1987A can be best reproduced
by models with a mass of
$M_{\rm ns,b}\simeq 1.5\,...\,1.6\,M_{\odot}$. The reference 
model of the simulations presented here was therefore chosen to
have $M_{\rm ns, b} = 1.53\,M_{\odot}$ and an initial
gravitational mass of $M_{\rm ns,g}^{\rm i} = 1.49\,M_{\odot}$.
The electron number fraction had a maximum at the center of
the star where $Y_{e, {\rm c}}^{\rm i} \simeq 0.29$.
The central temperature and density at the beginning of the
simulations were $T_{\rm c}^{\rm i} = 24\,{\rm MeV}$ and
$\rho_{\rm c}^{\rm i} = m\times 0.485\,{\rm fm}^{-3}\cong 
8\times 10^{14}\,{\rm g\,cm}^{-3} \simeq 3.25\rho_{\rm nuc}$,
respectively, with $\rho_{\rm nuc}$ being the nuclear matter
density and $m$ the common nucleon mass.

The computations were done with a general relativistic stellar
evolution code \cite{keil} employing equations of state developed
for the case $T = 0$ by Glendenning \cite{glen} and
extended to finite temperatures by including thermal corrections 
to the energy density and pressure of both stellar plasma and
neutrinos \cite{keil}. Two different equations of state were
employed. As the standard case, we used a hyperon equation of 
state (case~2 in \cite{glen}; EOS~B in \cite{keil})
which takes into account $n$, $p$, $e^{\pm}$, and $\mu^-$ in the
nuclear medium, and hyperons ($\Lambda$, $\Sigma\,...$) and
$\Delta$-resonances as additional hadronic states
at densities above about $2\rho_{\rm nuc}$.
Due to the formation of these additional baryonic degrees of
freedom, the hyperon equation of state is ``softer'' at very
high densities than a non-hyperonic equation of state involving
only $n$, $p$, $e^{\pm}$ and $\mu^-$. An equation of state of
the latter category (case~5 in \cite{glen}, EOS~A in \cite{keil}) 
was used for comparative computations in the work presented here.

The equations of state employed here do not predict the occurrence
of pions or pion condensates in the supernova core. The 
possible importance of pionic excitations was pointed out
in a paper by Mayle, Tavani, and Wilson \cite{mayleetal}. The
presence of a significant number of negative thermal pions influences
the axion emission from forming neutron stars by producing axions via 
pion-axion conversion processes $\pi^-+p \to n+a$ (Sect.~\ref{pion} and
\cite{mayle}). In order to estimate the impact of the latter on the 
protoneutron star cooling, we computed a number of models using the
rather crude assumption that one $\pi^-$ per nucleon is present at
densities beyond about two times nuclear matter density. 
The local abundance $Y_{\pi^-} = n_{\pi^-}/n_{\rm b}$ 
was simply prescribed by $Y_{\pi^-} = Y_{N}\cdot
\left\{1 + \tanh\left[2(\rho-\rho_{\rm nuc})/
\rho_{\rm nuc}\right]\right\}/2$ 
with $\rho_{\rm nuc} = m\times 0.15\,{\rm fm}^{-3}$. This
prescription is ad hoc and effects of pions on the nuclear 
equation of state were not taken into account at all.
Also, we ignored the temperature dependence of the abundance
of thermal pions.
Our choice of $Y_{\pi^-}$ of order unity was motivated by the results
of Ref.~\cite{mayleetal}, where an abundance of $Y_{\pi^-}\lesssim 0.6$
(and a total pion abundance of $Y_{\pi}\lesssim 0.8$) was obtained for
the conditions of temperature and density in a neutron star model at
$3\,{\rm s}$ after core bounce. The possibility of a pion condensate,
however, was excluded for the supernova conditions and the pion
dispersion relation assumed in Ref.~\cite{mayleetal}. Since
our whole treatment of the pion case was very approximate only,
we also did not make any attempt to include the contributions 
of pions to the neutrino scattering opacity of the 
nuclear medium (see also \cite{mayleetal}).

As for the neutrino transport, equilibrium
diffusion was assumed for all types of neutrinos ($\nu_e$,
$\bar\nu_e$, $\nu_{\mu}$, $\bar\nu_{\mu}$, $\nu_{\tau}$, 
$\bar\nu_{\tau}$) which is a good approximation, because the hot
matter of the protoneutron star is very opaque against neutrinos
by neutral-current neutrino-nucleon scatterings, $\nu+N \to \nu+N$, 
and charged-current $\beta$-reactions, $\nu_e+n\to e^- +p$ and
$\bar\nu_e + p\to e^+ + n$. For details of the technical 
implementation and the ``standard'' description of
neutrino-matter interactions, see \cite{keil}. A possible 
reduction of the neutrino opacity by many-body correlation
effects, e.g., a suppression of the $\nu N$ scattering cross
section by rapid spin fluctuations due to frequent nucleon-nucleon
collisions (\cite{R1,R2,R3,keiletal,Sigl} and 
Sect.~\ref{manyb}), was included only in a comparative model run
to reveal the change of the axion mass limit. We want to
emphasize here that the central density of our reference 
neutron star model with $M_{\rm ns, b} = 1.53\,M_{\odot}$
is always less than $8.8\times 10^{14}\,{\rm g\,cm}^{-3}$.
Only in a relatively small, central part with a mass 
$\lesssim 0.5\,M_{\odot}$ does the density become higher than 
about $2\rho_{\rm nuc}$ and hyperonization sets in. 
But even at the center of the star the hyperon abundance
$Y_{\rm hyp} = n_{\rm hyp}/n_{\rm b}$ never rises above 
$Y_{\rm hyp} \simeq 0.25$ (see Fig.~7 in \cite{glen}).
Therefore we consider the disregard of modifications of the
neutrino opacity due to the presence of hyperons \cite{prakash}
as acceptable for the models discussed in this work, and the
assumption that the neutrino opacity is produced mainly by 
interactions of neutrinos with $n$ and $p$ should yield a 
sufficiently accurate description.

In order to evaluate our protoneutron star cooling models
for the prediced neutrino signals in the KII and 
IMB detectors, we folded the computed spectral $\bar\nu_e$ 
number flux with the detector efficiency functions and the
cross section for $\bar\nu_e$ absorption on protons. A 
distance to SN~1987A of $D = 50\,{\rm kpc}$ was assumed. 
For details of the evaluation procedure, see Appendix~C of
\cite{keil}.

\section{Axion Emission Rates}

\subsection{Nucleon-Nucleon Axion Bremsstrahlung}

As long as the widths of the nucleon states are small compared to
the temperature, one can evaluate the matrix elements for
nucleon-pair bremsstrahlung of axions
by using free nucleon states. A one-pion exchange (OPE)
potential is likely to be an adequate starting-point
for describing the two-nucleon interaction. As a
function of the four-momentum transfer
between the nucleons, $k=(k_0,{\bf k})$, it can be written
as~\cite{Friman}
\begin{equation}
  V_{\rm OPE}({\bf k},{\hbox{\boldmath $\sigma$}}_1,
  {\hbox{\boldmath $\sigma$}}_2)=-\left({f\over m_\pi}\right)^2
  {\left({\hbox{\boldmath $\sigma$}}_1\cdot{\bf k}\right)
  \left({\hbox{\boldmath $\sigma$}}_2\cdot{\bf k}\right)
  \over k^2+m^2_\pi}\left({\hbox{\boldmath
  $\tau$}}_1\cdot{\hbox{\boldmath $\tau$}}_2\right)\,.\label{OPE}
\end{equation}
Here, $f\simeq1$ is the pion-nucleon coupling constant,
$m_\pi$ is the pion mass, and ${\hbox{\boldmath
$\sigma$}}_j$ and ${\hbox{\boldmath $\tau$}}_j$ are spin and
isospin operators for the two nucleons, respectively ($j=1,2$).
The resulting matrix element ${\cal M}$ was first calculated in
Ref.~\cite{Iwamoto} for degenerate nucleons and later
in~\cite{Brinkmann} for arbitrary nucleon degeneracy. The
lowest-order energy-loss rate per unit
volume due to axion emission, $Q^{(1)}_a$, is then given by the
phase-space integral
\begin{eqnarray}
  Q^{(1)}_a&=&\int{d^3{\bf k}_a\over2\omega(2\pi)^3}\prod_{j=1}^4
  {d^3{\bf p}_j\over2E_j(2\pi)^3}\,\omega f_1f_2(1-f_3)(1-f_4)
  \label{phspace}\\
  &&\times S\sum_{\rm spins}
  \vert{\cal M}\vert^2(2\pi)^4\delta^4(p_1+p_2-p_3-p_4-k_a)
  \,,\nonumber
\end{eqnarray}
where $p_j=(E_j,{\bf p}_j)$ are the four-momenta of the
initial-state ($j=1,2$) and final-state ($j=3,4$) nucleons,
and $k_a=(\omega,{\bf k}_a)$ is the axion four-momentum.
Furthermore, $f_j$ is the nucleon occupation number in state
$p_j$ ($j=1,\cdots,4$), and $S$ is the usual symmetry factor.

In Ref.~\cite{Brinkmann} the 15-dimensional integration in
Eq.~(\ref{phspace}) was performed exploiting the fact that
$\vert{\cal M}\vert^2$ varies only slightly in the range where the
integrand contributes most. Neglecting this variation
induces an error of less
than a factor 2. Introducing the thermal average
\begin{equation}
\xi\equiv3\left\langle\left[{({\bf p}_2-{\bf p}_4)\cdot
({\bf p}_2-{\bf p}_3)\over\vert{\bf p}_2-{\bf p}_4\vert
\vert{\bf p}_2-{\bf p}_3\vert}\right]^2\right\rangle
\label{thaverage}
\end{equation}
which can be shown to take the values
0 and 1.0845 in the limits of degenerate and
non-degenerate nucleons, respectively, the result
can be written as
\begin{eqnarray}
  Q^{(1)}_a&=&64\left({f\over m_\pi}\right)^4m^{2.5}T^{6.5}\Biggl[
  (1-\xi/3)g^2_{an}I(y_n,y_n)+(1-\xi/3)g^2_{ap}I(y_p,y_p)
  \nonumber\\
  &&+{4(15-2\xi)\over9}\left({g^2_{an}+g^2_{ap}\over2}\right)
  I(y_n,y_p)+{4(6-4\xi)\over9}\left({g_{an}+g_{ap}\over2}\right)^2
  I(y_n,y_p)\Biggr]\,.\label{Qanum}
\end{eqnarray}
Here, $y_i$ is the dimensionless non-relativistic
version of the nucleon chemical potential $\mu_i$:
$y_i=(\mu_i-m_i)/T$,
and the dimensionless function $I(y_1,y_2)$ can be fitted to
within 25\% accuracy by the analytic expression~\cite{Brinkmann}
\begin{eqnarray}
  I(y_1,y_2)&\simeq&\biggl[2.39\times10^5\left(e^{-y_1-y_2}+
  0.25e^{-y_1}+0.25e^{-y_2}\right)+1.73\times10^4(1+|\bar
  y|)^{-1/2}\nonumber\\
  &&+6.92\times10^4(1+|\bar y|)^{-3/2}+1.73\times10^4
  (1+|\bar y|)^{-5/2}\biggr]^{-1}\,,\label{Ifit}
\end{eqnarray}
with $\bar y=(y_1+y_2)/2$.

\subsection{Pion-Axion Conversion\label{pion}}

If pions or kaons are present in a supernova core, additional
processes such as $\pi^-+p\to n+a$ can contribute to
axion emission. Since it is uncertain whether a pion condensate
can form in the hot protoneutron star~\cite{mayleetal}, we
consider only thermal pions here. The corresponding lowest-order
perturbative 
energy emission rate per unit volume was found to be~\cite{mayle}
\begin{equation}
  Q_a^{\pi^-}=\frac{30f^2{\bar g}^2_{aN} T^3}{\pi m^2m_\pi^2}
  \,n_{\pi^-}n_p\,,\label{piminus}
\end{equation}
where ${\bar g}_{aN}$ is a combination of axion-proton and axion-neutron
couplings,
\begin{equation}
{\bar g}^2_{aN}={1\over 2}\left(g^2_{ap}+g^2_{an}\right) - 
                {1\over 3}\,g_{ap}g_{an} \, ,
                \label{gpi}
\end{equation}
and $n_{\pi^-}$
and $n_p$ are the number densities of $\pi^-$ and protons. If the 
pion abundance is comparable to the nucleon abundance, energy loss
by pion-axion conversion
will dominate over nucleon-pair bremsstrahlung of axions by
more than a factor of 10~\cite{mayle} in the perturbative approximation,
i.e., when saturation due to many-body effects in the dense medium
(see Sect.~\ref{manyb}) are ignored. We have considered the axion
mass bounds resulting from the lowest-order pion-axion emission 
rate as well. Due to the clear dominance of the axion emission from 
pion conversion processes in the perturbative approximation, we
neglected the production of axions by
nucleon-nucleon bremsstrahlung when $\pi^-$ were assumed to
be present in the cooling neutron star.

\subsection{Many-Body Effects\label{manyb}}

Recently the possibility was
discussed~\cite{R1,R2,R3,Sigl} that near nuclear densities
many-body effects might suppress substantially
the actual energy-loss rate in axions
compared to the lowest-order result Eq.~(\ref{Qanum}). 
A suppression is likely to occur also for the rate of 
Eq.~(\ref{piminus}). In order to discuss this issue and to compare
different approaches, it is convenient to reformulate the
loss rate in terms of the structure function formalism adopted
in these references. In the limit of non-relativistic nucleons,
axions couple exclusively to the dynamical spin-density structure
function (SSF), as can be seen from Eq.~(\ref{Lagran}). Its
long-wavelength limit is defined as
\begin{equation}
  S_\sigma(\omega)=\lim_{{\bf k}\to0}{4\over3n_b}
  \int_{-\infty}^{+\infty}
  dte^{i\omega t}\left\langle{\hbox{\boldmath $\sigma$}}(t,{\bf k})
  \cdot{\hbox{\boldmath $\sigma$}}(0,-{\bf k})
  \right\rangle\,.\label{sdef}
\end{equation}
Here, ${\hbox{\boldmath $\sigma$}}(t,{\bf k})=V^{-1}\int d^3{\bf r}
e^{-i{\bf k}\cdot{\bf r}}{\hbox{\boldmath $\sigma$}}(t,{\bf r})$
(with $V$ the normalization volume)
is the Fourier transform of the local nucleon spin-density
operator ${\hbox{\boldmath $\sigma$}}(t,{\bf r})$, $n_b$ is the
baryon density, and $(\omega,{\bf k})$ is the four-momentum transfer
to the medium. The expectation value $\langle\cdots\rangle$ in
Eq.~(\ref{sdef}) is taken over a thermal ensemble and the states
involved are normalized to $V$. For a single species of nucleons whose
coupling to axions is given by $C_i$ 
[see Eqs.~(\ref{Lagran})--(\ref{coupling})]
$Q_a$ can then be written as~\cite{R3,Sigl}
\begin{equation}
  Q_a={C_i^2n_b\over(4\pi)^2f_a^2}\int_0^\infty d\omega\,\omega^4
  e^{-\omega/T}S_\sigma(\omega)\,.\label{Qastruc}
\end{equation}

For two nucleon species one can absorb the constant $C^2_i$ into
the definition of $S_\sigma$ by multiplying
${\hbox{\boldmath $\sigma$}}(t,{\bf k})$ in Eq.~(\ref{sdef}) by
$\left[1+\tau_3(t,{\bf k})\right]C_p/2+\left[1-\tau_3(t,{\bf
k})\right]C_n/2$. Here, $\tau_3(t,{\bf k})$ is the third
component of the isospin operator ${\hbox{\boldmath
$\tau$}}(t,{\bf k})$ which is defined analogously to
${\hbox{\boldmath $\sigma$}}(t,{\bf k})$. For the qualitative
discussion in the rest of this section, it is sufficient to focus
on the case of a single nucleon species if not stated otherwise.

It is obvious from Eq.~(\ref{sdef}) that only
interactions which do not conserve the total nucleon spin can
lead to non-vanishing values of $S_\sigma(\omega)$ at
$\omega\neq0$.
As can be seen easily, the OPE potential Eq.~(\ref{OPE}) has this
property. For the following, it is convenient to
introduce the lowest-order effective spin fluctuation rate,
\begin{equation}
  \Gamma^{(1)}_\sigma\equiv{1\over3\pi}\left({f\over m_\pi}\right)^4
  m^2n_b\left\langle
  v\right\rangle\simeq32\,{\rm MeV}\rho_{14}T^{1/2}_{10}
  \,,\label{Gsigma}
\end{equation}
where $\left\langle v\right\rangle$ is the average relative
velocity between two nucleons, $\rho_{14}$ is the mass density
in units of
$10^{14}{\rm gcm}^{-3}$, and $T_{10}=T/10\,{\rm MeV}$. In the limit
of non-degenerate nucleons,
the lowest-order contribution to the SSF takes the
form~\cite{R2}
\begin{equation}
  S^{(1)}_\sigma(\omega)={K\Gamma^{(1)}_\sigma\over\omega^2}
  \,s_0(\omega/T)\,,\label{Ss1}
\end{equation}
where $K=12\pi^{-1/2}\left(T/m\right)^{1/2}/\left\langle
v\right\rangle\simeq2.7$~\footnote[1]{The relation between our
$\Gamma^{(1)}_\sigma$ defined in Eq.~(\ref{Gsigma}) and the quantity
$\Gamma_A$ used in Ref.~\cite{R2} is
$\Gamma_A=K\Gamma^{(1)}_\sigma$.}. The dimensionless bounded
function $s_0(\omega/T)$ has been given in Ref.~\cite{R2}.
Adopting Eq.~(\ref{Ss1}) in Eq.~(\ref{Qastruc}) leads to a rate $Q_a$
which, for the case of two nucleon species, coincides with
Eq.~(\ref{Qanum}) in the
non-degenerate limit, $y_n,y_p\ll-1$. Note that, according to
Eq.~(\ref{Ifit}), $I(y_1,y_2)\propto e^{y_1+y_2}\propto
n_1n_2T^{-3}$ in this limit, where $n_i$ is the number density
of species $i$.

However, Eq.~(\ref{Ss1}) is unphysical in the limit of both
small and large energy transfers, which can be seen as follows.
First, one can derive the sum rule,
\begin{equation}
  \int_{-\infty}^{+\infty}{d\omega\over2\pi}\,S_\sigma(\omega)
  =1+{4\over3n_bV}\left\langle\sum_{i\neq j}{\hbox{\boldmath
  $\sigma$}}_i\cdot{\hbox{\boldmath $\sigma$}}_j\right\rangle
  \,,\label{sum1}
\end{equation}
where the sum over all nucleon
pairs accounts for possible spin correlations among different
nucleons. At least at low densities, these correlations can be
neglected compared to the first term. In any case, the
finiteness of the integral in Eq.~(\ref{sum1}) clearly shows
that the infrared singularity in Eq.~(\ref{Ss1}) is
unphysical. In fact, higher-order effects are expected to
regularize this singularity at low energy transfers. As a first
qualitative guess, it was suggested~\cite{R2} to substitute
$\omega^{-2}$ by $(\omega^2+a^2\Gamma^2_\sigma)^{-1}$, where $a$
is a dimensionless number of order unity which can be chosen to
satisfy the sum rule Eq.~(\ref{sum1}).

Secondly, one of us (G.S.)~\cite{Sigl} recently derived the
analog of what is usually called the f sum rule,
\begin{equation}
  \int_{-\infty}^{+\infty}{d\omega\over2\pi}\,\omega S_\sigma(\omega)
  =-{4\over n_bV}\langle H_T\rangle\equiv{2\Gamma_\sigma\over\pi}
  \,,\label{sum2}
\end{equation}
where $H_T={1\over2}\sum_{i\neq j}V^{T}_{ij}$
is the ``tensor component'' of the nucleon interaction
Hamiltonian, defined as the tensor component
$V^T_{ij}$ of the two-nucleon interaction potential~\cite{BD},
summed over all pairs. It is this
component which violates local nucleon spin conservation and,
according to Eq.~(\ref{sum2}), determines the width of the SSF
via the effective spin fluctuation rate
\begin{equation}
\Gamma_\sigma\equiv{-2\pi\left\langle H_T\right\rangle\over n_bV} \ .
\label{gamma}
\end{equation}
$\Gamma^{(1)}_\sigma$ defined in
Eq.~(\ref{Gsigma}) can be shown to be the first-order approximation to
$\Gamma_\sigma$ in a dilute medium. It turns out that in the case
of two nucleon species the f sum, Eq.~(\ref{sum2}), diverges
when Eq.~(\ref{Ss1}) is substituted for the SSF~\cite{Sigl}.
This can be traced back to the unphysical behavior of the
dipole-like OPE potential Eq.~(\ref{OPE}) at small distances.
More realistic potentials which account for hard-core repulsion
lead to a fall off of $S_\sigma$ at high $\omega$ which is
stronger than $\omega^{-2}$, thus assuring f sum
integrability.

A plausible modification of the SSF is given by
\begin{equation}
  S_\sigma(\omega)={K\Gamma_\sigma\over\omega^2+a^2\Gamma^2_\sigma}
  \,s(\omega/T)\,,\label{Ss}
\end{equation}
where the continuous and bounded function $s(\omega/T)$ is even
and satisfies $s(0)=1$. This expression has the right limiting
behavior in the classical regime which obtains for $\omega\ll
T$~\cite{Raffelt}. For $\omega\lesssim\Gamma_\sigma$, multiple collisions
become important and lead to a suppression of
$S_\sigma(\omega)$ compared to the lowest-order approximation
Eq.~(\ref{Ss1}), which is known as the Landau-Pomeranchuk-Migdal
effect~\cite{LPM,KV} and ensures integrability of the sum
rule Eq.~(\ref{sum1}). In the classical bremsstrahlung limit of
hard collisions one would have $s(x\equiv\omega/T)\equiv1$. Quantum
corrections require that $s(x)\to0$ for $x\to\infty$
sufficiently fast, as can be seen from the f sum rule
Eq.~(\ref{sum2}). The exact shape of $s(x)$ depends on the
nucleon interaction potential. In particular, the high $x$ behavior is
governed by the small-distance regime which is usually dominated
by a hard-core repulsion.

Modifications of $s(\omega/T)$ at
$\omega\gtrsim T$ do not have a big influence on the loss rate
Eq.~(\ref{Qastruc}), because of the exponential factor. The finite-width
modification is also unimportant as long as $\Gamma_\sigma\lesssim
T$, i.e., in the dilute medium. In this regime, we have
$\Gamma_\sigma\simeq\Gamma^{(1)}_\sigma$ and $Q_a\simeq
Q^{(1)}_a\propto n_b\Gamma^{(1)}_\sigma T^3$ [see Eq.~(\ref{Qanum})].
For $\Gamma_\sigma\gg T$, in contrast, $Q_a/(n_bT^4)$
would start to decrease with
increasing $\Gamma_\sigma$~\cite{Sigl}. This would be the case if
$\Gamma_\sigma\simeq\Gamma^{(1)}_\sigma$ up to the highest
densities [see Eq.~(\ref{Gsigma})]. However, in Ref.~\cite{Sigl}
it was argued that $\Gamma_\sigma$ is likely to saturate at some
maximum value $\Gamma^{\rm max}_\sigma\lesssim150\,{\rm MeV}$. This also
implies that neutrino opacities are suppressed by less than
$\simeq50\%$ compared to the lowest-order opacities, ensuring
consistency of the observed and simulated cooling time scales for 
SN~1987A~\cite{keiletal}. For $\Gamma_\sigma\lesssim150\,{\rm MeV}$,
$Q_a\simeq Q^{(1)}_a$ is a good approximation. For a
first improvement to account for saturation of $\Gamma_\sigma$, we
therefore set 
\begin{equation}
  Q_a = Q^{(1)}_a\,{\rm Min}\left[1,{\Gamma^{\rm max}_\sigma \over
        \Gamma^{(1)}_\sigma}\right] \quad {\rm with} \quad
  \Gamma^{\rm max}_\sigma \equiv 2\pi W 
  \label{Qsat}
\end{equation}
in the numerical simulations, with $Q^{(1)}_a$ taken from 
Eq.~(\ref{Qanum}). The maximum value of the spin
fluctuation rate, $\Gamma^{\rm max}_\sigma$, is coined in terms of
the average interaction energy $W$ of a nucleon in the nuclear
medium. This definition is motivated by Eq.~(\ref{gamma}) and 
implies the relation $W = -\left\langle H_T\right\rangle/(n_bV)$.

We have made no systematic attempt to include the effects of a 
saturation of nucleon spin fluctuations due to many-body interactions
in case of the pion-axion conversion process.
However, irrespective of whether the process involves
real pions or virtual pions from a bystander nucleon,
axions are emitted by the fluctuating nucleon spin whose fluctuation
rate $\Gamma_\sigma$ is expected to saturate according to 
$\Gamma^{\rm max}_\sigma$ of Eq.~(\ref{Qsat}).
Inclusion of saturation effects is therefore likely to leave axion
bounds considerably less sensitive to the abundance of pions 
than suggested by the lowest-order energy-loss rate of 
Eq.~(\ref{piminus}).

\section{Axions and Protoneutron Star Cooling}

\subsection{Models}

In this section we discuss our protoneutron star cooling 
calculations which were performed with a systematic variation of the 
axion-nucleon couplings and with different input physics in the
neutron star modeling. The considered
axion-nucleon coupling constants $g_{ai}$ are of the order
of $10^{-10}$ and thus axion opacity plays no role \cite{burres}.

In our first sequence of cooling models, the loss of energy by
nucleon-pair axion brems\-strah\-lung [Eqs.~(\ref{Qanum}) and
(\ref{Ifit})] was investigated in dependence of the two parameters
$g_{ap}$ and $x\equiv g_{an}/g_{ap}$ which were chosen from the
intervals $0 \le g_{ap}/10^{-10} \le 10$ and $-3 \le x \le 3$,
respectively. Suppression or saturation of the axion emission
rates was not taken into account. The nucleon degeneracy parameter
$\xi$ of Eq.~(\ref{thaverage}) was set to $0.5$, but comparative 
calculations with $\xi = 1.0845$ (nondegenerate nucleons) and
$\xi = 0$ (very degenerate nucleons) revealed only a very weak
dependence of the results on the particular value of $\xi$. 

In a second sequence of models we repeated these cooling 
calculations with many-body effects taken into account according
to Eq.~(\ref{Qsat}) where $Q_a^{(1)}$ is from Eq.~(\ref{Qanum})
and $\Gamma_{\sigma}^{(1)}$ is defined in Eq.~(\ref{Gsigma}).
The average interaction energy per nucleon, $W$, was chosen to
be $W = 10\,{\rm MeV}$, corresponding to $\Gamma_\sigma^{\rm max}
\simeq 60\,{\rm MeV}$. A finite value of $W$ leads to a 
saturation of the axion emission rate for large values of 
$\Gamma_{\sigma}^{(1)}$, i.e., at high densities and/or high
temperatures. Our ``standard'' case without saturation is formally
recovered for $\Gamma_\sigma^{\rm max} \propto W \to \infty$.
We also performed comparative computations for the values 
$W = 5\,{\rm MeV}$ and $W = 20\,{\rm MeV}$. 

In a third set of models we considered the effects of the
presence of a large number of negative pions in the nuclear 
medium. Pions were included in an ad hoc manner as 
described in Sect.~\ref{pns-ph} by assuming $0.5< Y_{\pi^-}< 1$
for the number of $\pi^-$ per nucleon at densities above 
$\rho_{\rm nuc}$.
The rate of energy loss via the pion-axion conversion reaction
is given by Eq.~(\ref{piminus}). Because of the 10--50 times larger
rate compared to nucleon-nucleon axion bremsstrahlung, a range of
coupling parameters $0\le {\bar g}_{aN}/10^{-10}\le 2$ was explored.

In addition to these three sets of models and the variations of
$\xi$ and $W$ mentioned above, we replaced the hyperon equation
of state (EOS~B) by an ordinary $n$-$p$ equation of state (EOS~A) in
order to reveal the differences for the axion production during the
Kelvin-Helmholtz cooling of the protoneutron star when hyperons and
$\Delta$-resonances are not present at densities beyond
$2\rho_{\rm nuc}$.
Furthermore, the influence of a reduction of the neutrino opacity at
high densities due to rapid nucleon spin fluctuations \cite{R1,R2,R3,Sigl}
was explored. A suppression factor of 50\% was assumed, which is the 
maximum reduction still allowed by consistency between the SN~1987A
neutrino detections and the calculated neutrino signals for axionless
cooling models \cite{keiletal}.
Finally, the sensitivity of the derived axion mass limits on the mass
of the cooling neutron star was tested by performing cooling 
calculations not only for our standard protoneutron star model with
$M_{\rm ns, b} = 1.53\,M_{\odot}$, but also for models with baryonic 
masses $M_{\rm ns, b} = 1.30\,M_{\odot}$, $1.40\,M_{\odot}$, 
$1.65\,M_{\odot}$, and $1.75\,M_{\odot}$. Like most of the other 
comparative computations, the simulations with different neutron star 
masses were done with the reference values $g_{ap}=1.5\times 10^{-10}$
and $x = 0$.

\subsection{Protoneutron Star Evolution without and with Axion Emission}

For sufficiently large axion-nucleon coupling the cooling of
protoneutron stars is significantly affected by the production
of freely escaping axions. We compare here the results for
cooling calculations of the $1.53\,M_{\odot}$ neutron star,
once without axions ($g_{ap} = g_{an} = 0$)
and another time with the representative choice for the axion
coupling constants of
$g_{ap} = 1.5\times 10^{-10}$ and $x = g_{an}/g_{ap} = 0$.
The energy loss rate due to axion emission was prescribed according to
Eq.~(\ref{Qanum}), i.e., suppression or saturation of the axion
emission at high densities and temperatures was not taken into account.
The temperature evolution for the axionless case is shown in
Fig.~\ref{F2}, the corresponding information for the case with axions
is given in Fig.~\ref{F3}. 

The initial temperature profile ($t = 0$) corresponds to the 
situation a few ten milliseconds after core bounce. The 
temperature shows a flat hump between the (baryonic) mass 
coordinates of $M(r) \equiv N_b(r) m \sim 0.3\,M_{\odot}$ and
$\sim 0.9\,M_{\odot}$ ($N_b(r)$ is the total baryon number inside 
radius $r$).
The temperature in these intermediate layers of the protoneutron
star is higher than near the center because of the heating caused by 
the forming supernova shock. Moreover, as the deleptonization of the 
protoneutron star progresses, electron degeneracy energy which is
released in the process $e^-+p\to n+\nu_e$, is not completely
radiated away from the surface of the star by the emission of all
kinds of neutrinos. Instead, due to downscattering and multiple
absorption and reemission of diffusing neutrinos, a part of this 
energy stays in the star and leads to heating of the gas. Therefore
the temperature in the interior of the star 
rises during the first $\sim 7\,{\rm s}$ of the 
evolution (Fig.~\ref{F2}) and the temperature peak advances inward to the
center of the star, following the motion of the layer where
most of the lepton (electron) loss occurs. At the time when the 
temperature maximum has reached the center of the star (after
$\sim 7\,{\rm s}$), the chemical potential
$\mu_{\nu_e} = \mu_e + \mu_p - \mu_n$ of electron neutrinos
has dropped to its final equilibrium value $\mu_{\nu_e} = 0$
that corresponds to deleptonized, neutronized conditions. Now
the star begins to cool down
essentially coherently by the continuing energy loss due to the 
emission of neutrino-antineutrino pairs created by
thermal processes. After about $38\,{\rm s}$ the temperatures are
below $\sim 3\,{\rm MeV}$ in the whole star.

If axions are produced by nucleon-nucleon
bremsstrahlung in significant amounts, two 
consequences follow for the cooling. On the one hand, the maximum
temperature in the star reaches only about $35\,{\rm MeV}$
at the mass coordinate $M(r)\sim 0.7\,M_{\odot}$ (Fig.~\ref{F3}).
This has to be compared with the peak temperature of
$\sim 48\,{\rm MeV}$ realized at the center of the
star for the case without axion emission (Fig.~\ref{F2}). 
On the other hand, the temperature
starts to drop right from the beginning and already after 
$18\,{\rm s}$ the whole star has cooled down to a temperature of 
less than $3\,{\rm MeV}$ everywhere.
Obviously, axions are very efficient in transporting away the
heat produced after the conversion of degenerate electrons into
neutrinos.

In Fig.~\ref{F4} the local energy emission rate
due to nucleon-nucleon axion bremsstrahlung is plotted against
the enclosed baryonic rest mass $M(r)$ for different times.
The axion emission rate per 
baryon peaks where $\rho T^{3.5}$ has a local maximum and is
largest during the first few seconds of the evolution. At later
times the temperatures in the star have dropped appreciably and
the energy loss in axions is greatly reduced. This is underlined
by Fig.~\ref{F5} which shows the axion luminosity as a function of time
in comparison with the combined luminosities of $\nu_e$ and 
$\bar\nu_e$ (denoted by ``$L_{\nu_e}$''), the combined luminosities
of $\nu_{\mu}$, $\bar\nu_{\mu}$, $\nu_{\tau}$ and $\bar\nu_{\tau}$
(labeled by ``$L_{\nu_{\mu}}$''), and the total luminosity for
all neutrinos ($L_{\nu}$). The axion luminosity decreases by
an order of magnitude within $5\,{\rm s}$, a time after which
$L_{\nu}$ has dropped by only a factor of 3--4. Note that the
luminosities in Fig.~\ref{F5} include reductions due to gravitational 
redshift and time dilation for an observer at rest at infinity.
Since neutrinos diffuse out through the star and decouple from the
stellar matter near the surface of the protoneutron star, their
luminosities are corrected for the gravitational redshift at the 
stellar surface which is typically 25\% (of the energy
measured at infinity). Axions, instead,
leave the star from the deep interior where most of the axion
production takes place (see Fig.~\ref{F4}) and their typical redshift is 
40--50\%. Time dilation (i.e., redshift of the inverse time
interval or frequency) is accounted for by the same factors.
Therefore the amount of energy transported away from the star
by axions relative to the energy emitted in neutrinos is larger 
than suggested by the integration of the luminosities
depicted in Fig.~\ref{F5}.

\subsection{Parameter Studies}

Increasing the axion-proton and axion-neutron couplings 
$g_{ap}$ and $g_{an}$, respectively, leads to a decrease of
the energy radiated in neutrinos. Fig.~\ref{F6} displays the total energies
lost by the $1.53\,M_{\odot}$ protoneutron star in axions ($E_a^0$) 
and in neutrinos ($E_{\nu}^0$) as functions of the coupling
parameter $g_{ap}$. The nucleon-pair axion bremsstrahlung is again
described by Eq.~(\ref{Qanum}).
The symbols mark computed models and are connected by cubic
spline interpolation. The different lines correspond to the
values $x = g_{an}/g_{ap} = 0\,, \pm 0.5\,, \pm 1.0$. The axion
energy is larger for larger absolute values of $x$ and a slight
difference between the results for positive and negative $x$ 
(with the values of $E_a^0$ being a bit larger for positive $x$)
reflects the asymmetry of the last term in Eq.~(\ref{Qanum}) against
changes of the sign of $g_{an}$. In Fig.~\ref{F6}, both $E_a^0$ and 
$E_{\nu}^0$ are the energies as measured by a locally inertial
observer at rest at the surface of the protoneutron star. 
The axion energy is therefore corrected for the gravitational 
redshift between the layers of the axion
production and the neutron star surface.

Fig.~\ref{F7} shows the number of $\bar\nu_e$ absorption events 
predicted for
the KII detector, $N_{\rm KII}$, and Fig.~\ref{F8} the time interval 
$t_{\rm KII}$ within which 90\% of these events are registered.
Figs.~\ref{F9} and~\ref{F10} give the corresponding information for the
IMB detector. Because the fraction of the gravitational binding
energy emitted in neutrinos decreases when more energy is carried 
away by axions (Fig.~\ref{F6}), all quantities displayed in
Figs.~\ref{F7}--\ref{F10}
exhibit a rapid reduction with increasing $g_{ap}$. This reduction
is stronger for larger $|x|$. For an
axion-proton coupling constant between $1\times 10^{-10}$ and
$2\times 10^{-10}$, the expected numbers of detector events and 
the detection times are reduced to about half of the values 
for the axionless case. While the calculated 11
neutrino events within $\sim 15\,{s}$ for the $1.53\,M_{\odot}$
protoneutron star are in reasonable agreement with the KII
measurement of SN~1987A neutrinos (11 neutrinos in $12.5\,{\rm s}$), 
the $\sim 6$ predicted 
IMB events within $\sim 9\,{\rm s}$ are clearly on the low
side of the detection rate of SN~1987A neutrinos in IMB
(8 neutrinos in $5.6\,{\rm s}$). This trend of
our results is also present for supernova models of other groups 
and reflects the marginal consistency observed between the IMB and KII
neutrino data: For the $\bar\nu_e$ emission characteristics
(spectrum and luminosity) deduced from the KII measurement, one
would expect a much smaller number of IMB events, or, inversely,
$\bar\nu_e$ emission that has caused the 8 events in IMB should have
produced a much larger signal in the KII detector 
(see, e.g., \cite{suzuki}).

If axion emission is absent, the expected event numbers $N_{\rm KII}$
and $N_{\rm IMB}$ as well as the detection times $t_{\rm KII}$
and $t_{\rm IMB}$ are slowly increasing with the mass of the 
protoneutron star \cite{keiletal}. Between 
$M_{\rm ns,b}=1.30\,M_{\odot}$ and $M_{\rm ns,b}=1.75\,M_{\odot}$,
$N_{\rm KII}$ and $N_{\rm IMB}$ roughly double, while 
$t_{\rm KII}$ increases by a factor of 2.5 and $t_{\rm IMB}$
by about 60\%, as can be seen in Table~1 of \cite{keiletal}. 
When axions are produced in the stars by
nucleon-pair bremsstrahlung [Eq.~(\ref{Qanum})] with
axion-proton coupling $g_{ap} = 1.5\times 10^{-10}$ and
axion-neutron coupling $g_{an} = 0$ (i.e., $x = 0$), the results
for different neutron star masses are obtained as shown in Fig.~\ref{F11}. All
four quantities reveal a very weak dependence on the baryonic mass
of the neutron star. For the used values of the axion-nucleon
coupling constants, we find $N_{\rm KII}\simeq 5$, 
$N_{\rm IMB} \simeq 3$, $t_{\rm KII} \simeq 6\,{\rm s}$, and
$t_{\rm IMB}\simeq 3.8\,{\rm s}$ in all cases. Independent of the
protoneutron star mass, these results are about half of
the values found for the case of the $1.53\,M_{\odot}$ star
without axion emission. The same inert behavior is also observed for the
average energies of the electron antineutrinos captured in the 
two detectors: $\left\langle\epsilon_{\rm KII}\right\rangle
\simeq 23.7\,{\rm MeV}$ and $\left\langle\epsilon_{\rm IMB}\
\right\rangle\simeq 34.5\,{\rm MeV}$.
In more massive protoneutron stars the 
temperature and density become higher; this causes enhanced
emission of axions so that a smaller fraction of the released
gravitational binding energy ends up in neutrinos. The weak 
variation of the expected neutrino signals suggests that the 
additional gravitational energy release associated with a
larger neutron star mass is essentially completely carried
away by axions.

Our study with varied parameters $g_{ap}$ and $x$, based on 
the $1.53\,M_{\odot}$ protoneutron star model, was
repeated for the case that the axion emission rate saturates
at high densities and temperatures (see Sect.~\ref{manyb}).
Employing Eq.~(\ref{Qsat}) with $Q_a^{(1)}$ from Eq.~(\ref{Qanum})
(with $\xi = 0.5$) and $\Gamma_{\sigma}^{(1)}$ from 
Eq.~(\ref{Gsigma}), the axion emission rate reaches its saturation
level when $\rho_{14}T_{10}^{1/2}\gtrsim 2W(10\,{\rm MeV})^{-1}$.
For an average interaction energy per nucleon of $W = 10\,{\rm MeV}$
we found that the corresponding reduction of the axion production 
increases the predicted number of neutrino events $N_{\rm KII}$ and 
$N_{\rm IMB}$ by roughly 40\% compared to the number 
of expected events for 
the ``naive'' (unsaturated) case for which formally $W \to \infty$.
This trend turned out to be even somewhat stronger for the 
detection times $t_{\rm KII}$ and $t_{\rm IMB}$ where the increase
was about 60\%. For $W = 20\,{\rm MeV}$ the increase was
$\sim 20$\% in the event numbers and $\sim 30$\% 
in the detection times, and for $W = 5\,{\rm MeV}$ the event
numbers and detection times rose by 60\% and 80\%, respectively.

Taking into account the possible existence of a large number
of negative pions at high densities in the simple
way described in Sect.~\ref{pns-ph} and using 
the energy loss rate due to pion-axion conversion as given
in Eq.~(\ref{piminus}), a reduction of the predicted detector
response ($N_{\rm KII}$, $N_{\rm IMB}$, $t_{\rm KII}$ and
$t_{\rm IMB}$) to 75\% of the values for the axionless case
occurred for ${\bar g}_{aN}\simeq 0.15\times 10^{-10}$. A reduction 
by 50\% was seen when ${\bar g}_{aN} = (0.30\,...\,0.40)\times 10^{-10}$,
and only 25\% of the neutrino events were measured in roughly
four times shorter time when ${\bar g}_{aN}$ was between 
$0.75\times 10^{-10}$ and $1.0\times 10^{-10}$. Of course,
the neutrino emission characteristics of the simulated
protoneutron stars vary with time. Therefore, the detection rates
in KII and IMB are not constant, and, correspondingly, the values of 
the axion-pion coupling ${\bar g}_{aN}$ which lead to a certain
reduction are somewhat different for the two experiments and 
also for the event numbers and detection times. 

Our results exhibit an extremely weak dependence on the choice of
the nucleon degeneracy parameter $\xi$ of Eq.~(\ref{thaverage}).
Changing $\xi$ from 0 (degenerate nucleons) to $1.0845$ 
(nondegenerate nucleons) leads to an increase of $N_{\rm KII}$
by $\sim 0.5$ events and to a detection time $t_{\rm KII}$
that is $\sim 0.5\,{\rm s}$ longer. For the IMB detector the
corresponding numbers are $\sim 0.25$ events and $\sim 0.25\,{\rm s}$,
respectively. A larger value for $\xi$ reduces the energy-loss
rate by axion emission [Eq.~(\ref{Qanum})], but the particular
choice of $\xi$ is irrelevant at the level of accuracy implied
by the sparseness of the neutrino data of SN~1987A. 

Replacing the hyperonic
EOS~B by the $n$-$p$ EOS~A in our $1.53\,M_{\odot}$ protoneutron
star model leads to essentially no change in the predicted 
neutrino signals. Hyperonization plays an important role only
in stars with baryonic masses $\gtrsim 1.70\,M_{\odot}$ where the
density in a larger part of the star is high enough to favor the
production of hadronic states other than $n$ and $p$ 
(see \cite{keil} and compare also
models S4BH\_0 and S4AH\_0 of Table~1 in \cite{keiletal}).

A suppression of the neutrino opacities relative to their 
``standard'' values, e.g., by the many-body effects discussed
in Sect.~\ref{manyb}, can have an important impact on the 
predicted neutrino signal \cite{keiletal} and thus on the 
allowed range of axion-nucleon couplings deduced from the
SN~1987A neutrino data. As a test, we reduced the axial vector
contributions of neutral- and charged-current processes
by about 50\% (i.e., we chose $a = 0.5$,
see \cite{keiletal}) in the case of the $1.53\,M_{\odot}$
protoneutron star with axion emission prescribed 
according to Eq.~(\ref{Qanum})
and $g_{ap} = 1.5\times 10^{-10}$ and $x = 0$. The 
characteristics of the detector signals change in the same
way as found for the axionless case in \cite{keiletal}.
A 50\% reduction of the neutrino opacities leads to an increase 
of the predicted numbers of neutrino events $N_{\rm KII}$ 
and $N_{\rm IMB}$ by $\sim 2.2$, but the detection times
$t_{\rm KII}$ and $t_{\rm IMB}$ drop by about 25\%. The
mean energies of $\bar\nu_e$ registered by the detectors 
rise by roughly $2\,{\rm MeV}$. Since these changes are not
dramatic, it is clear that even
a 50\% reduction of the neutrino opacity in the protoneutron
star will not completely alter conclusions on the axion production
drawn from a comparison of theoretical neutrino signals with
the SN~1987A neutrino data. Although the increase of the
number of detector events in case of a lower neutrino opacity
can somehow compensate for the effects of axion emission,
the detection times shrink {\it both} by a reduction of
the neutrino cross sections {\it and} by the additional axion cooling 
of the star. Therefore we conclude that a value of the neutrino opacity
that is lower than the ``standard'' one can hardly lead to a 
restoration of the compatibility between the SN~1987A neutrino
signal and the signal predicted in case of strong axion emission.

\subsection{Excluded axion couplings and axion mass limits}

The cooling sequences of the $1.53\,M_{\odot}$ protoneutron star
model for varied axion-nucleon coupling parameters $g_{ap}$ and
$x = g_{an}/g_{ap}$ are used to construct exclusion curves in 
the $x$-$g_{ap}$-space. The sensitivity against changes 
of the axion couplings was found to be somewhat different for predicted
event numbers and detection time scales. Nevertheless, it is safe
to claim that consistency between the theoretical neutrino signal 
and the SN~1987A neutrino data requires that 
$N_{\rm KII, IMB}\gtrsim {1\over 2} N_{\rm KII, IMB}^{\rm s}$ and
$t_{\rm KII, IMB}\gtrsim {1\over 2} t_{\rm KII, IMB}^{\rm s}$ when
$N_{\rm KII, IMB}^{\rm s}$ and $t_{\rm KII, IMB}^{\rm s}$ are the
expected signal parameters for the axionless ``standard''
$1.53\,M_{\odot}$ protoneutron star model. According to this 
criterion, the bell-shaped curves in Fig.~\ref{F12} separate allowed
from forbidden regions. The lower curve corresponds to the case
where the energy loss in axions is described by Eq.~(\ref{Qanum}),
the upper curve represents the case where saturation of axion emission
is included [Eq.~(\ref{Qsat})] with an average interaction energy
per nucleon of $W = 10\,{\rm MeV}$. The values of the upper curve
can be obtained from those of the lower curve by scaling with a 
factor of about 1.9. 

On the hatched sides above the
curves the axion-nucleon couplings are so large that more
than about half of the gravitational binding energy of the neutron star
is emitted in axions and only less than half in neutrinos (compare
Fig.~\ref{F6}).
There is only a slight asymmetry between positive and negative values 
of $x$ which is caused by the minor asymmetry of the third term of
$Q_a^{(1)}$ in Eq.~(\ref{Qanum}) against changes of the sign of $g_{an}$.
Obviously, the contribution of the asymmetric term 
$\propto g_{an}g_{ap}$ is rather small. For this reason, the lower
curve of Fig.~\ref{F12} can be pretty well fitted by the relation
$g_{ap}^2 + 2g_{an}^2 \simeq (1.5\times 10^{-10})^2$ which 
means that the excluded values of the axion-nucleon coupling
constants lie outside of an ellipse with semiaxes $1.5\times 10^{-10}$
and $(1.5/\sqrt{2})\times 10^{-10}$ in the $g_{ap}$-$g_{an}$-plane.

The deduced limits for allowed values of the coupling constants 
$g_{ap}$ and $x = g_{an}/g_{ap}$ might be sensitive to the protoneutron 
star mass. This was tested by varying the baryonic mass of the
protoneutron star for $g_{ap} = 1.5\times 10^{-10}$ and $x = 0$.
In case of the $1.53\,M_{\odot}$ model this pair of values ($x$,$g_{ap}$)
lies on the lower exclusion curve of Fig.~\ref{F12}. The results of 
such a study are displayed in Fig.~\ref{F11}. The observed inertia
of the predicted neutrino signal against changes of
the mass of the axion-emitting star suggests that the exclusion
curves should not depend strongly on the
(unknown) exact mass of the neutron star born in SN~1987A.

Taking into account the possible existence of a large number of negative 
thermal pions in the protoneutron star in the simple way described in 
Sect.~\ref{pns-ph}, we find that the rapid energy loss by axions 
emitted in pion-axion conversion processes [Eq.~(\ref{piminus})]
excludes couplings of roughly
$\bar g_{aN}\gtrsim (0.30\,...\,0.40)\times 10^{-10}$. This assumes
a perturbative treatment of the emission processes and neglects
possible saturation effects. Since for 
typical conditions in the supernova core the perturbative energy emission 
rate by pion-axion conversion is approximately 10--20 times
larger than the lowest-order energy-loss rate for the nucleon-nucleon 
axion bremsstrahlung, the bound on the coupling constant
$\bar g_{aN}$ would be 3--5 times more stringent than the limit on
$g_{ap}$ in the perturbative case.

\begin{figure}[ht]
\centering\leavevmode
\epsfxsize=5.2in
\epsfbox{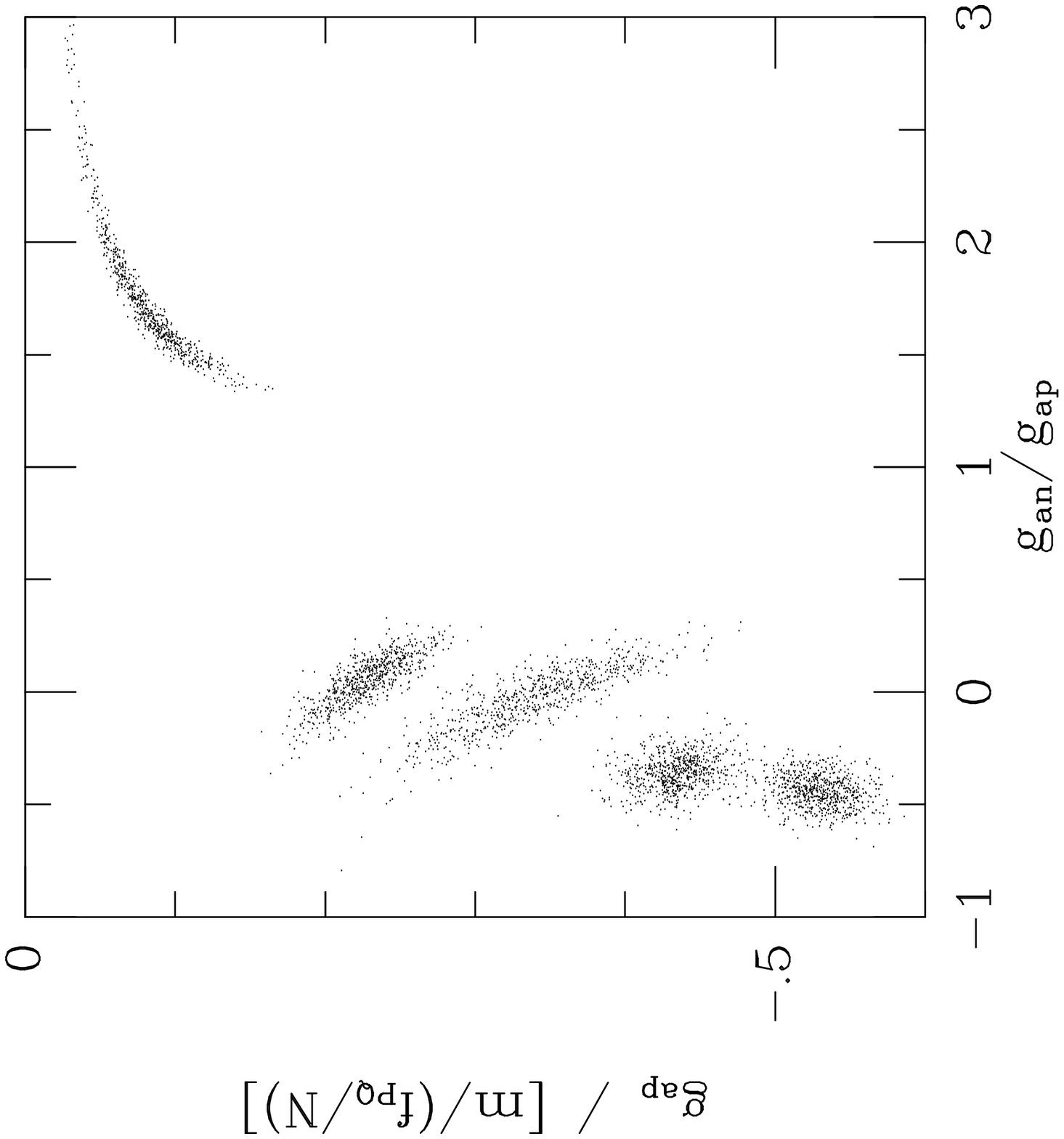}
\bigskip
\caption[...]{\baselineskip 14pt
Scatter plot of the axion-proton coupling and the
ratio of the axion-neutron to axion-proton coupling for different
axion models allowing for the uncertainties in $\Delta q$; from
the bottom left to the top right, $\beta = 0$, $\beta =27^\circ$,
KSVZ, $\beta = 54^\circ$, and $\beta = 81^\circ$.  Note
that the axion-neutron coupling is much smaller than the axion-proton
coupling for the KVSZ axion and for the DFSZ axion when $\beta \sim
45^\circ$.
\label{F1}}
\end{figure}

\begin{figure}[ht]
\centering\leavevmode
\epsfxsize=5.2in
\epsfbox{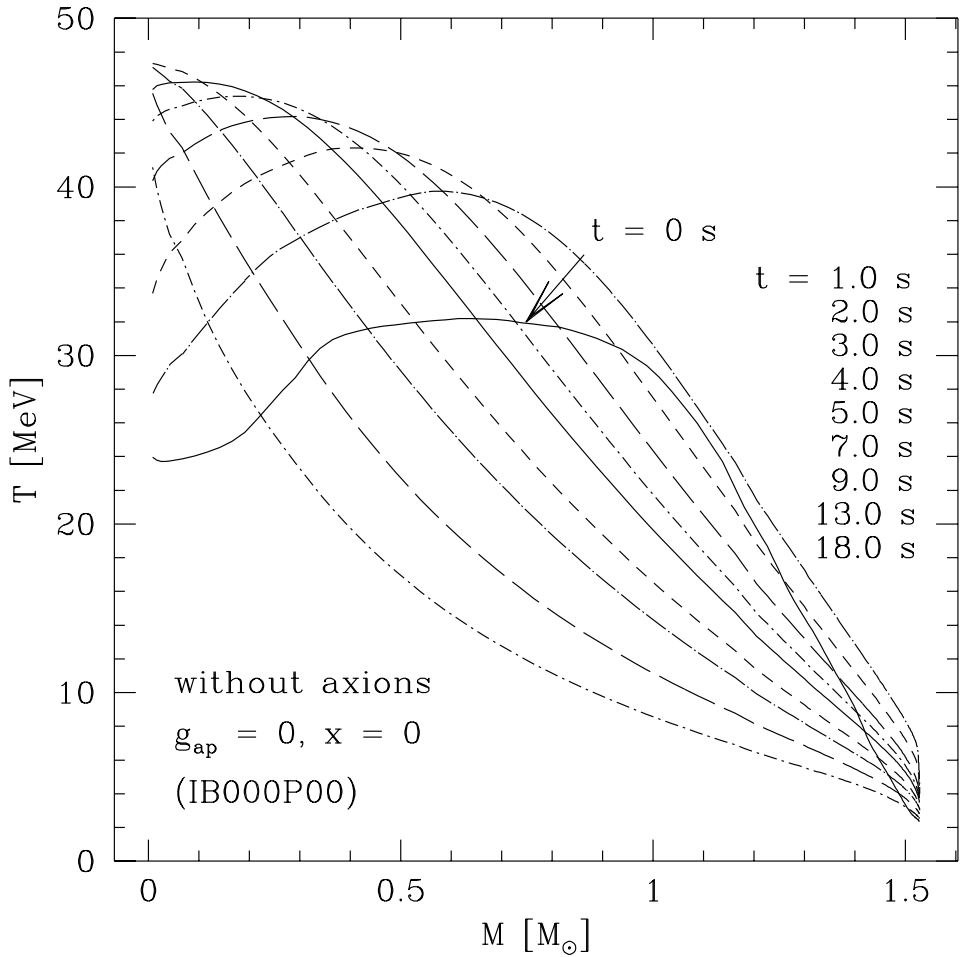}
\medskip
\caption[...]{
Temperature evolution of a $1.53\,M_{\odot}$ protoneutron star 
without axion emission. The temperature profiles are plotted
against the enclosed baryonic rest mass $M(r) = N_b(r)m$
(in solar masses; $m$ is the common nucleon mass) for 
different times from the start of the computation shortly after the
formation of the protoneutron star. For time $t = 0$ the curve is
labeled, the other times are listed according to the order of 
the corresponding curves (from top to bottom at $M = 1\,M_{\odot}$).
The interior of the star heats up first due to the conversion of
lepton degeneracy energy into thermal energy, but finally cools.
\label{F2}}
\end{figure}

\begin{figure}[ht]
\centering\leavevmode
\epsfxsize=5.2in
\epsfbox{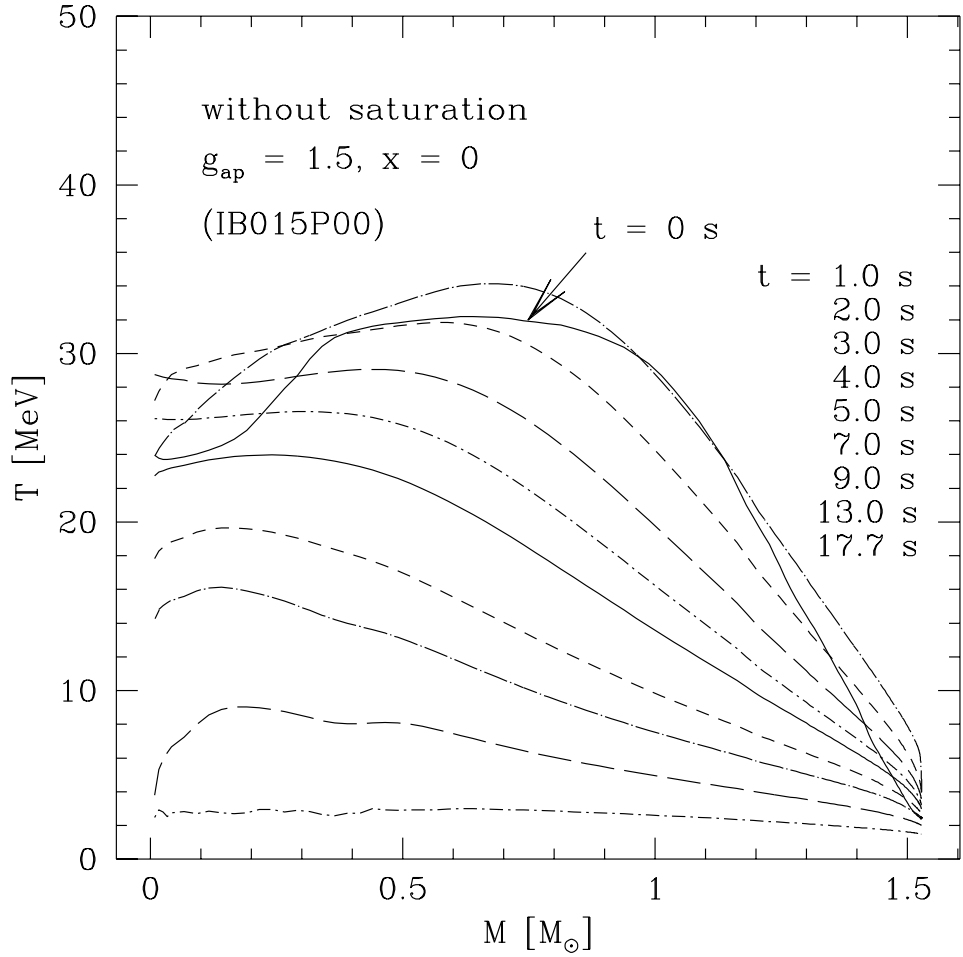}
\medskip
\caption[...]{
Temperature profiles of a $1.53\,M_{\odot}$ protoneutron star
vs.~enclosed baryonic mass for different times. Axion production
via nucleon-pair bremsstrahlung is included with axion coupling
constants $g_{ap} = 1.5\times 10^{-10}$ and $x = g_{an}/g_{ap} = 0$. 
The star cools much faster than in Fig.~\ref{F2}.
\label{F3}}
\end{figure}

\begin{figure}[ht]
\centering\leavevmode
\epsfxsize=5.2in
\epsfbox{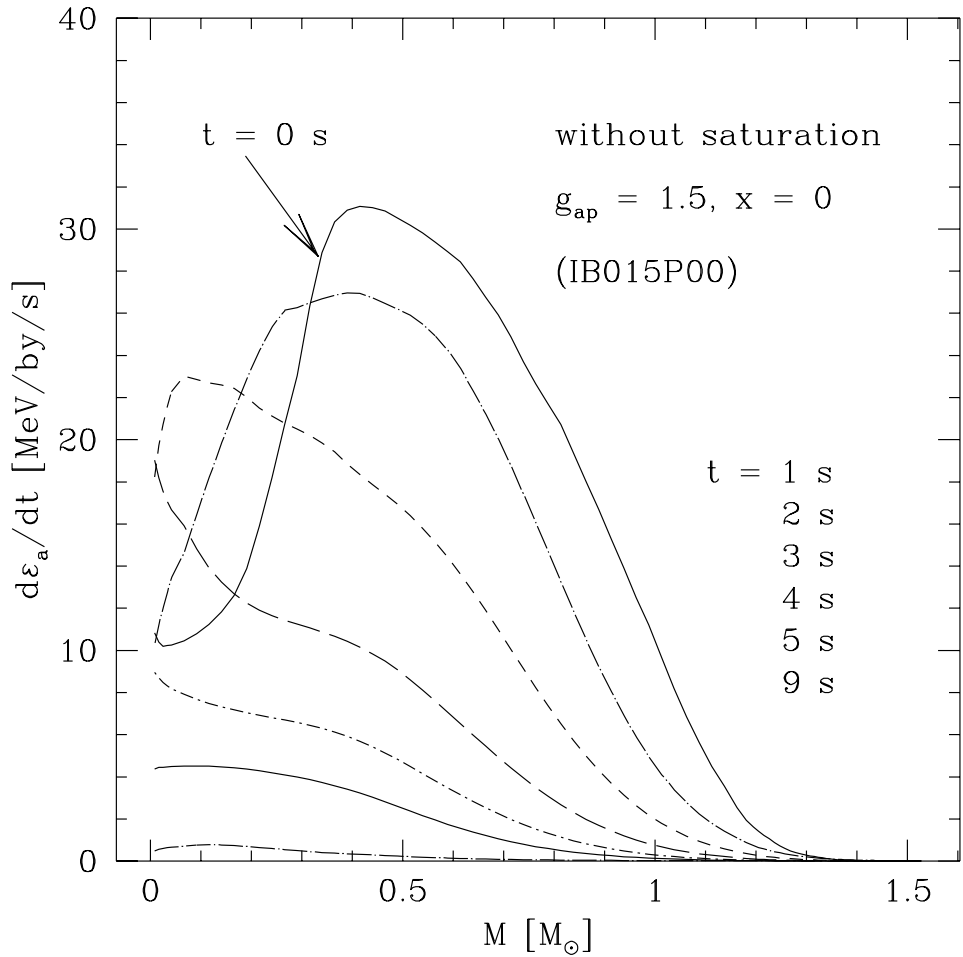}
\medskip
\caption[...]{
Local energy production rate per baryon due to nucleon-nucleon 
bremsstrahlung emission of axions [Eq.~(\ref{Qanum})]
with coupling constants $g_{ap} = 1.5\times 10^{-10}$ and
$x = g_{an}/g_{ap} = 0$. The profiles are plotted against the
baryonic mass coordinate of a $1.53\,M_{\odot}$ protoneutron 
star for different times after the start of the simulation
(the times are listed according to the order of the curves at 
$M = 0.5\,M_{\odot}$).
The energy production per baryon roughly peaks where the 
product $\rho T^{3.5}$ reaches a maximum.
\label{F4}}
\end{figure}

\begin{figure}[ht]
\centering\leavevmode
\epsfxsize=5.2in
\epsfbox{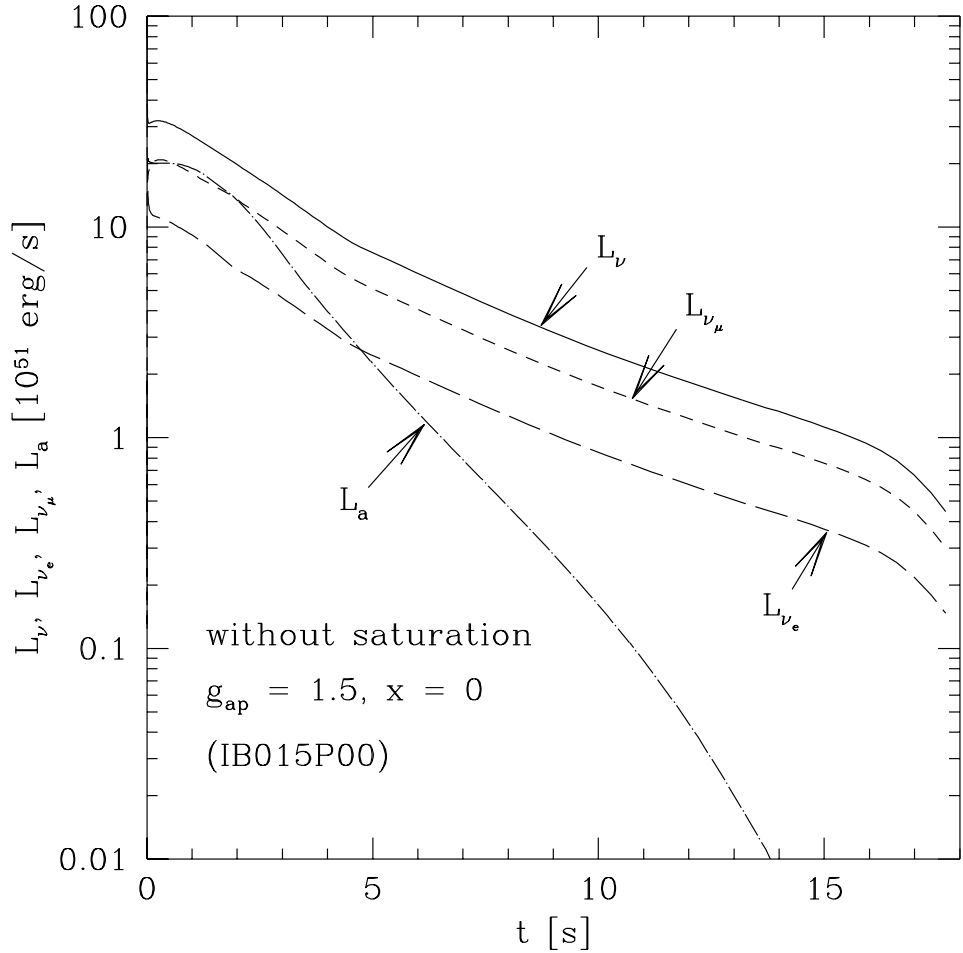}
\medskip
\caption[...]{
Electron neutrino plus electron antineutrino luminosity 
(labeled by $L_{\nu_e}$), sum of all luminosities of heavy lepton
neutrinos
(labeled by $L_{\nu_{\mu}}$), total neutrino luminosity $L_{\nu}$,
and axion luminosity $L_a$ as functions of time during the 
cooling of a $1.53\,M_{\odot}$ protoneutron star for axion-proton
coupling $g_{ap} = 1.5\times 10^{-10}$ and axion-neutron coupling
$g_{an} = 0$. The luminosities are redshifted as measured by an 
observer at rest at infinity. The axion luminosity drops by an
order of magnitude within $5\,{\rm s}$, whereas the neutrino emission
decreases less rapidly.
\label{F5}}
\end{figure}

\begin{figure}[ht]
\centering\leavevmode
\epsfxsize=5.2in
\epsfbox{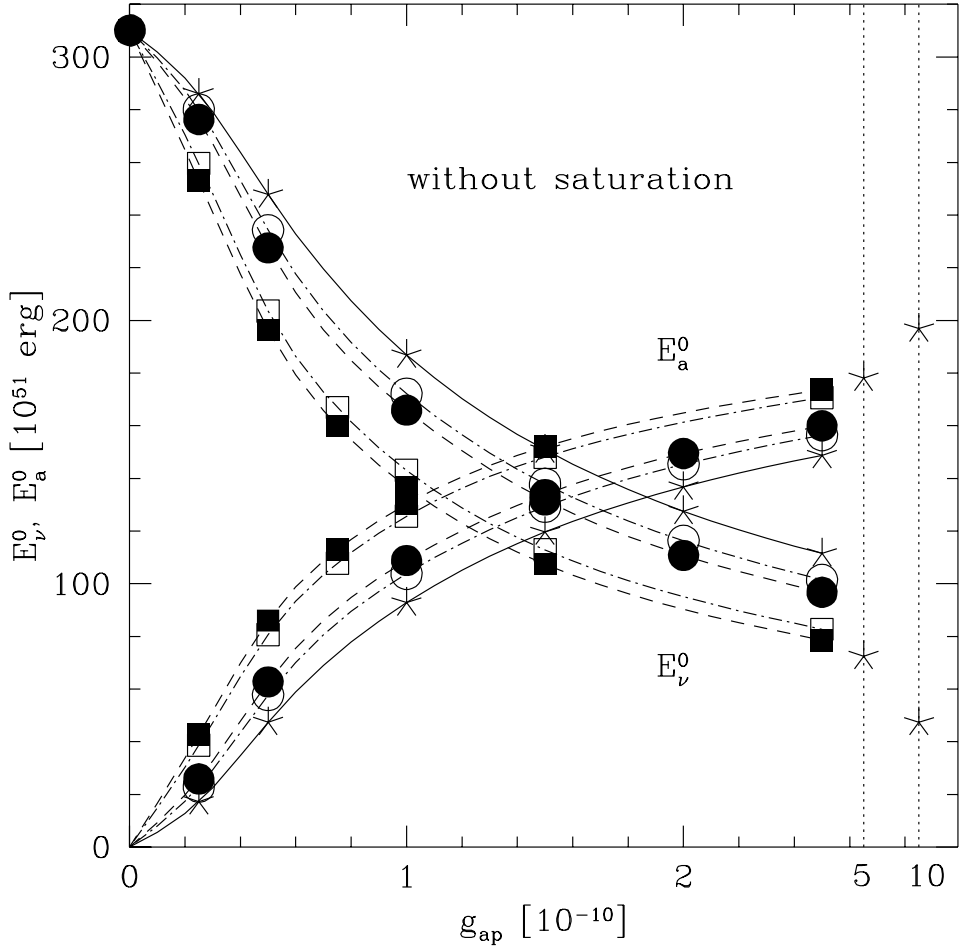}
\medskip
\caption[...]{
Energy loss of a $1.53\,M_{\odot}$ protoneutron star in neutrinos
($E_{\nu}^0$) and in axions ($E_a^0$) for different combinations
of values of the axion-proton coupling $g_{ap}$ and axion-neutron
coupling $g_{an}$ in Eq.~(\ref{Qanum}). 
The energies are given as measured by a locally
inertial observer at the surface of the neutron star. The symbols
mark computed models, the interpolation is done by cubic splines.
The curves through the asterisks correspond to the case
$x = g_{an}/g_{ap} = 0$, the circles to the cases $x = \pm 0.5$,
and the squares to $x = \pm 1.0$. Filled symbols mark positive,
open symbols negative values of $x$.
\label{F6}}
\end{figure}

\begin{figure}[ht]
\centering\leavevmode
\epsfxsize=5.2in
\epsfbox{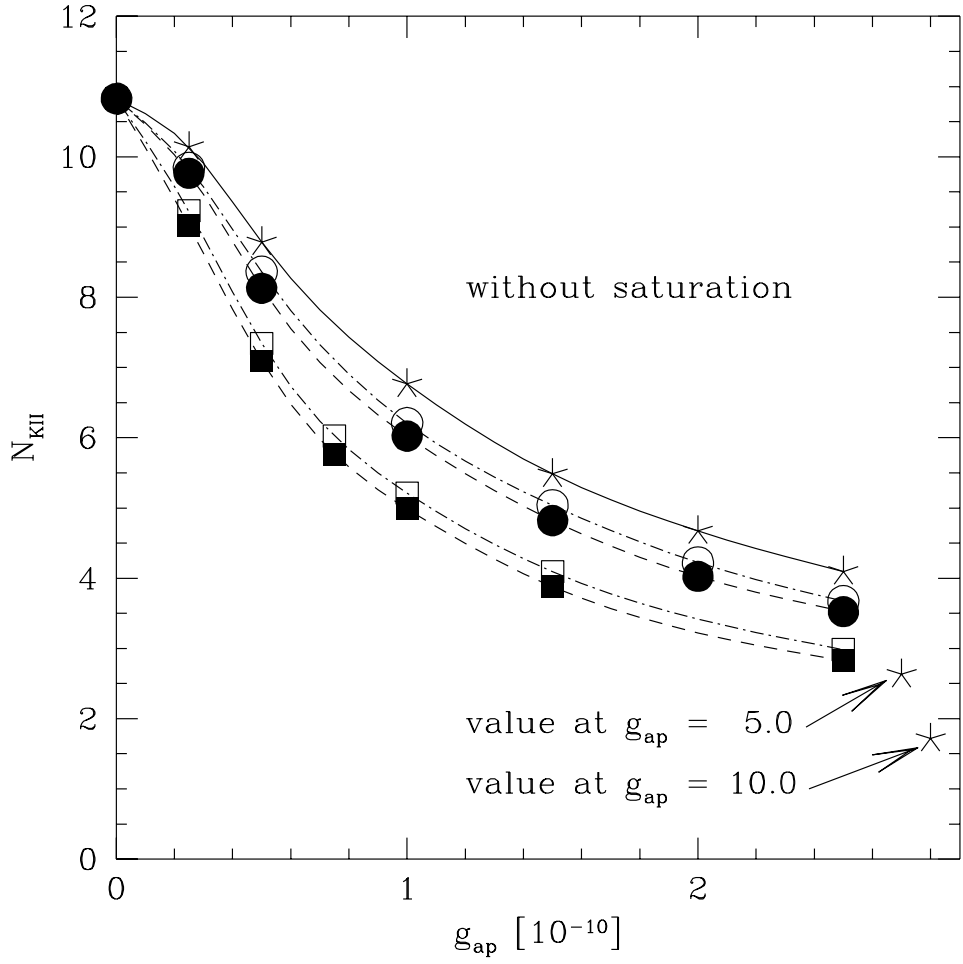}
\medskip
\caption[...]{
Prediced number of $\bar\nu_e$ absorption events in the KII
detector for a $1.53\,M_{\odot}$ protoneutron star as a function
of the axion-proton coupling $g_{ap}$. The symbols correspond
to computed models and the different curves represent different values
of $x = g_{an}/g_{ap}$ (see Fig.~\ref{F6}).
\label{F7}}
\end{figure}

\begin{figure}[ht]
\centering\leavevmode
\epsfxsize=5.2in
\epsfbox{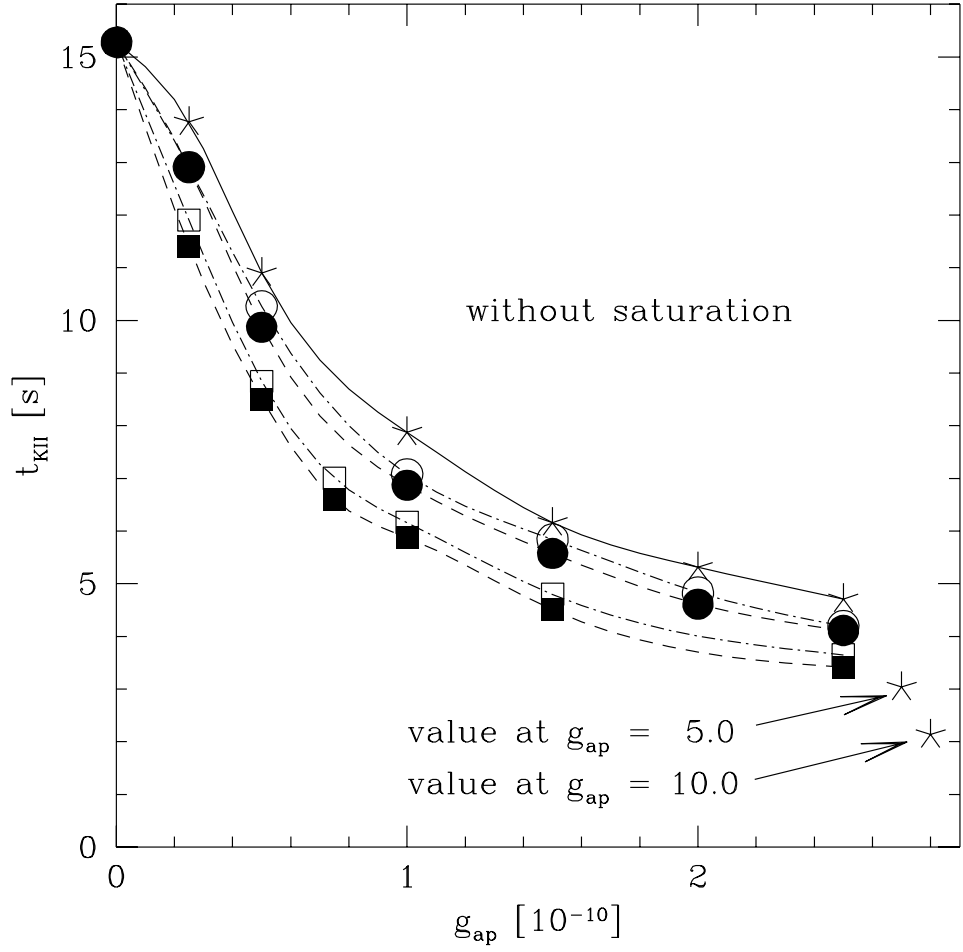}
\medskip
\caption[...]{
Time intervals $t_{\rm KII}$ where 90\% of the predicted
$\bar\nu_e$ capture events in the KII detector happen for a 
protoneutron star with a baryonic mass of $1.53\,M_{\odot}$
and for different combinations of values of $g_{ap}$ and
$x = g_{an}/g_{ap}$. The meaning of the symbols is the same as
in Fig.~\ref{F6}.
\label{F8}}
\end{figure}

\begin{figure}[ht]
\centering\leavevmode
\epsfxsize=5.2in
\epsfbox{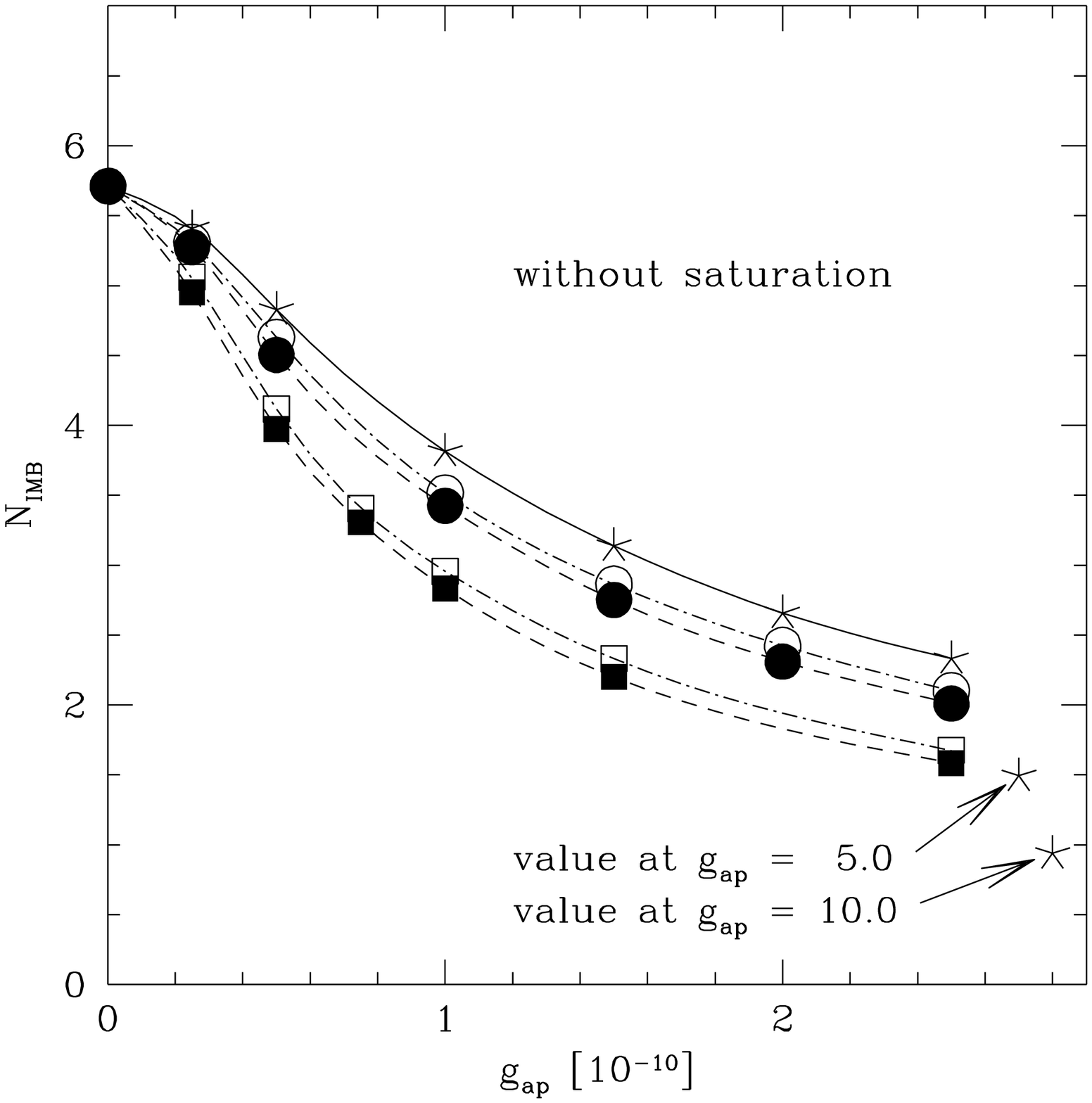}
\medskip
\caption[...]{
Same as Fig.~\ref{F7}, but for the IMB detector.
\label{F9}}
\end{figure}

\begin{figure}[ht]
\centering\leavevmode
\epsfxsize=5.2in
\epsfbox{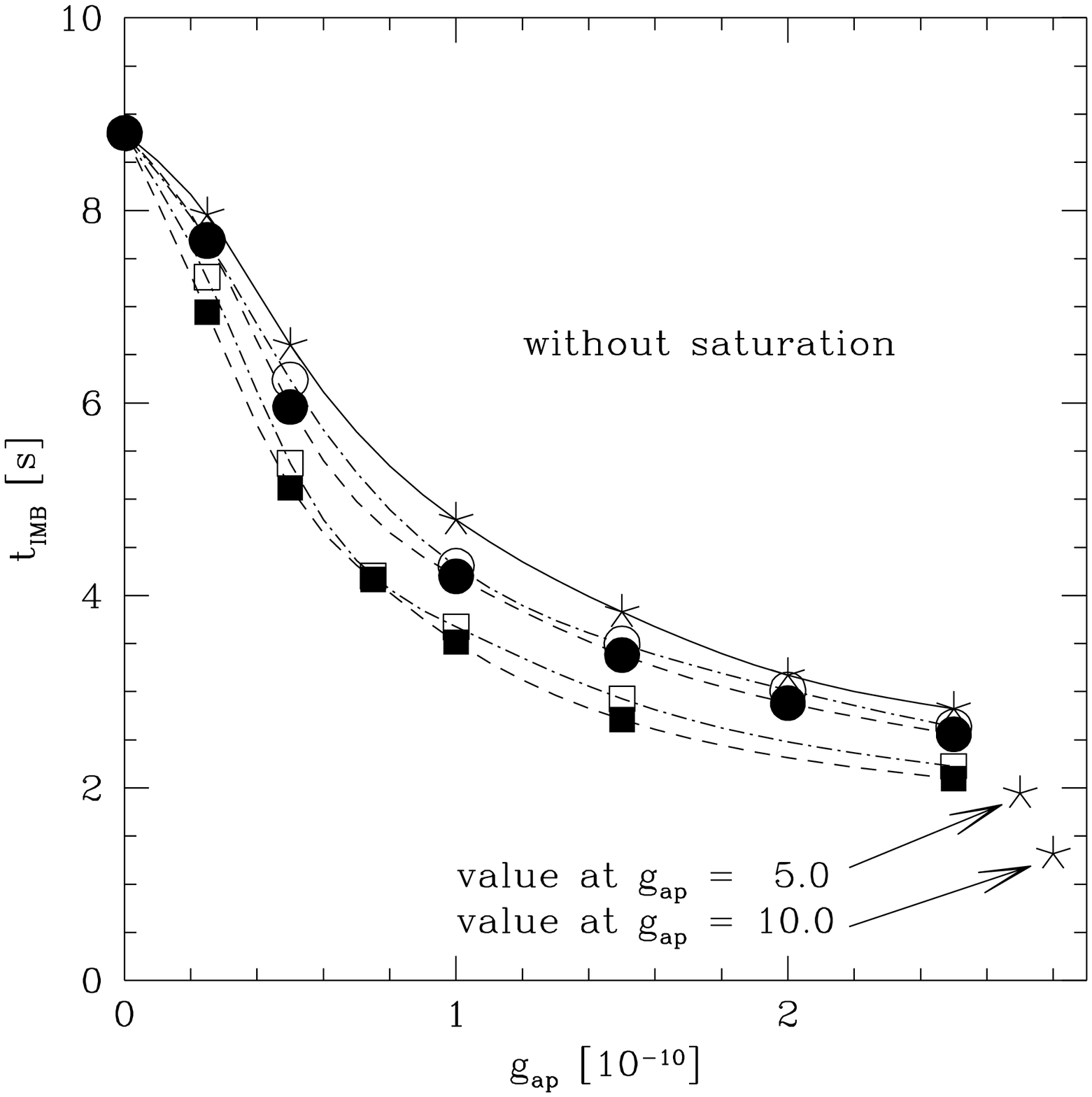}
\medskip
\caption[...]{
Same as Fig.~\ref{F8}, but for the IMB detector.
\label{F10}}
\end{figure}

\begin{figure}[ht]
\centering\leavevmode
\epsfxsize=5.2in
\epsfbox{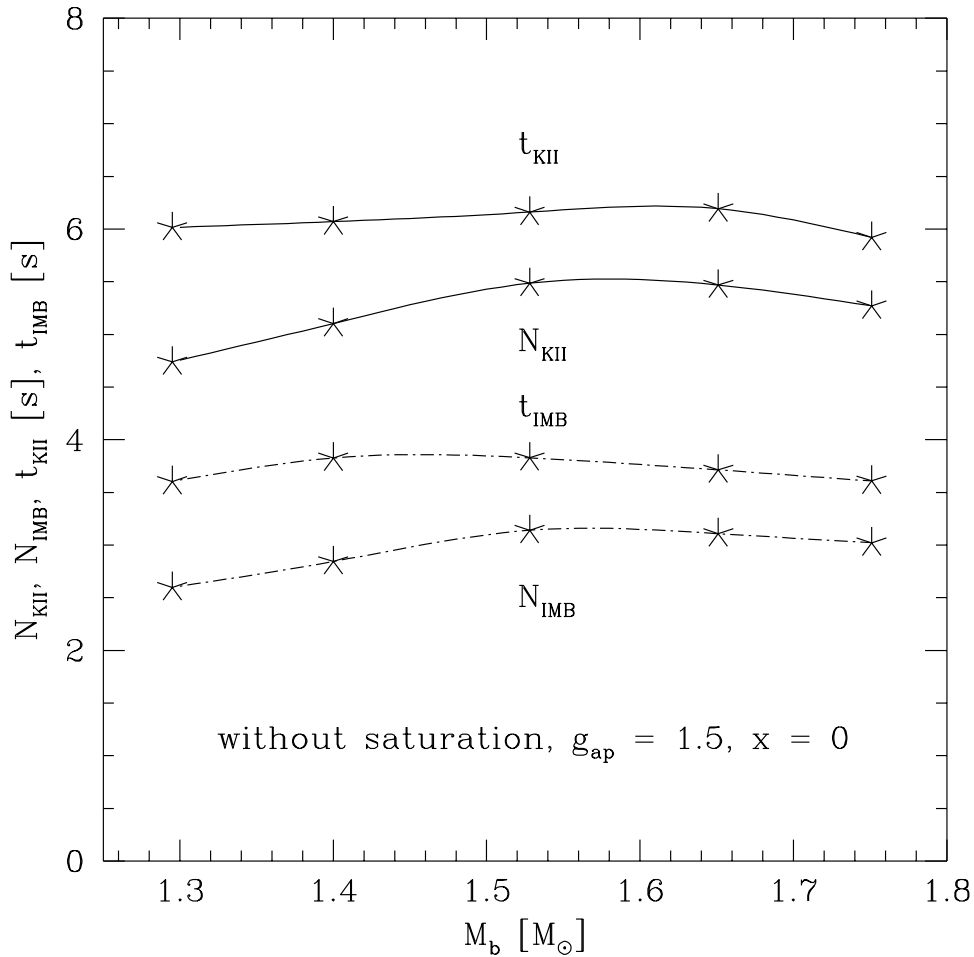}
\medskip
\caption[...]{
Numbers of predicted $\bar\nu_e$ absorption events in the KII
and IMB detector, $N_{\rm KII}$ and $N_{\rm IMB}$, respectively,
and corresponding time intervals $t_{\rm KII}$ and $t_{\rm IMB}$ 
where 90\% of these events happen. Five protoneutron star models 
with different baryonic masses, $M_{\rm ns,b} = 1.30\,M_{\odot}$,
$1.40\,M_{\odot}$, $1.53\,M_{\odot}$, $1.65\,M_{\odot}$,
and $1.75\,M_{\odot}$ were evolved with axion emission by
nucleon-nucleon bremsstrahlung [Eq.~(\ref{Qanum})] for 
$g_{ap} = 1.5\times 10^{-10}$ and $x = g_{an}/g_{ap} = 0$.
The expected measurements for event numbers and detection times
vary extremely weakly with $M_{\rm ns,b}$.
\label{F11}}
\end{figure}

\begin{figure}[ht]
\centering\leavevmode
\epsfxsize=5.2in
\epsfbox{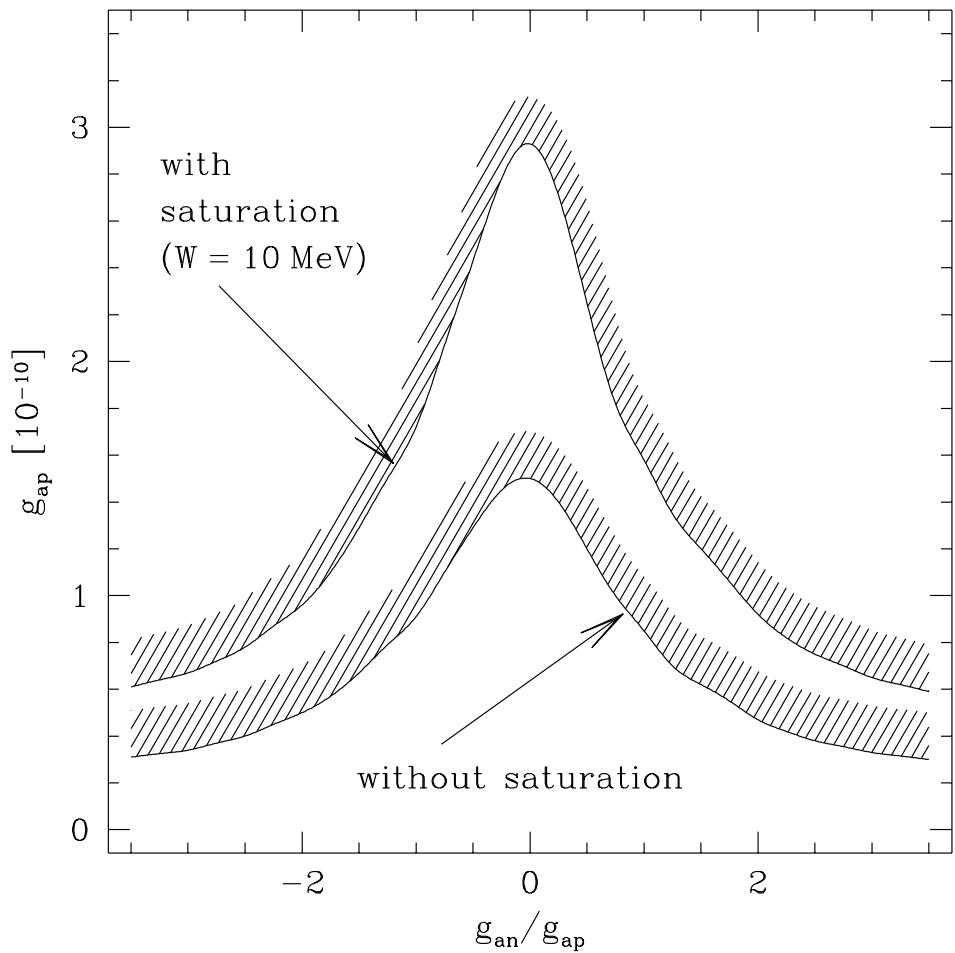}
\medskip
\caption[...]{
SN~1987A limit to the axion proton coupling
as a function of the ratio of the axion-neutron to axion-proton couplings,
with and without saturation due to many-body effects.  Saturation
(for an average interaction energy per nucleon of $W = 10\,{\rm MeV}$)
degrades the limit by a factor of about 1.9 independent of the
ratio of the axion-neutron to axion proton couplings. The combinations
of coupling parameter values above the bell-shaped curves are 
excluded.
\label{F12}}
\end{figure}

With these limits on the axion coupling parameters $g_{ap}$,
$g_{an}$ (or, equivalently, $x$), and $\bar g_{aN}$, upper limits on the 
axion mass can be derived from Eqs.~(\ref{defc}) and (\ref{gpi})
when use is made of the experimental values for the dimensionless
couplings $C_p$ and $C_n$ (see Sect.~\ref{axcoup}) collected in Fig.~\ref{F1},
and Eq.~(\ref{maxion}) is employed to relate the PQ scale (or 
axion decay constant) $f_a$ with the axion mass $m_a$. From 
Eqs.~(\ref{defc}) and (\ref{maxion}) one finds
\begin{eqnarray}
{m_a\over 10^{-3}\,{\rm eV}} \lesssim {0.66\over C_p}\,
{g_{ap}^{\rm max}\over 10^{-10}} \, ,
\end{eqnarray}
and from Eqs.~(\ref{gpi}) and (\ref{maxion}) one obtains
\begin{eqnarray}
{m_a\over 10^{-3}\,{\rm eV}} \lesssim {0.66\over C_p}\,
\left({1\over 2}+{x^2\over 2} - {x\over 3}\right)^{\! -1/2}\,
{\bar g_{aN}^{\rm max}\over 10^{-10}} \, .
\end{eqnarray}
Here $g_{ap}^{\rm max}$ and $\bar g_{aN}^{\rm max}$ are the
maximum values of the axion couplings allowed by the SN~1987A
neutrino detections, and $C_p = g_{ap}/[m_p/(f_{\rm PQ}/N)]$ and
$x$ correspond to the pair of coordinate values of a particular
chosen point in Fig.~\ref{F1}. Note that $g_{ap}^{\rm max}$ as a function
of $x$ describes the (lower) exclusion curve in Fig.~\ref{F12}.

\begin{figure}[ht]
\centering\leavevmode
\epsfxsize=5.2in
\epsfbox{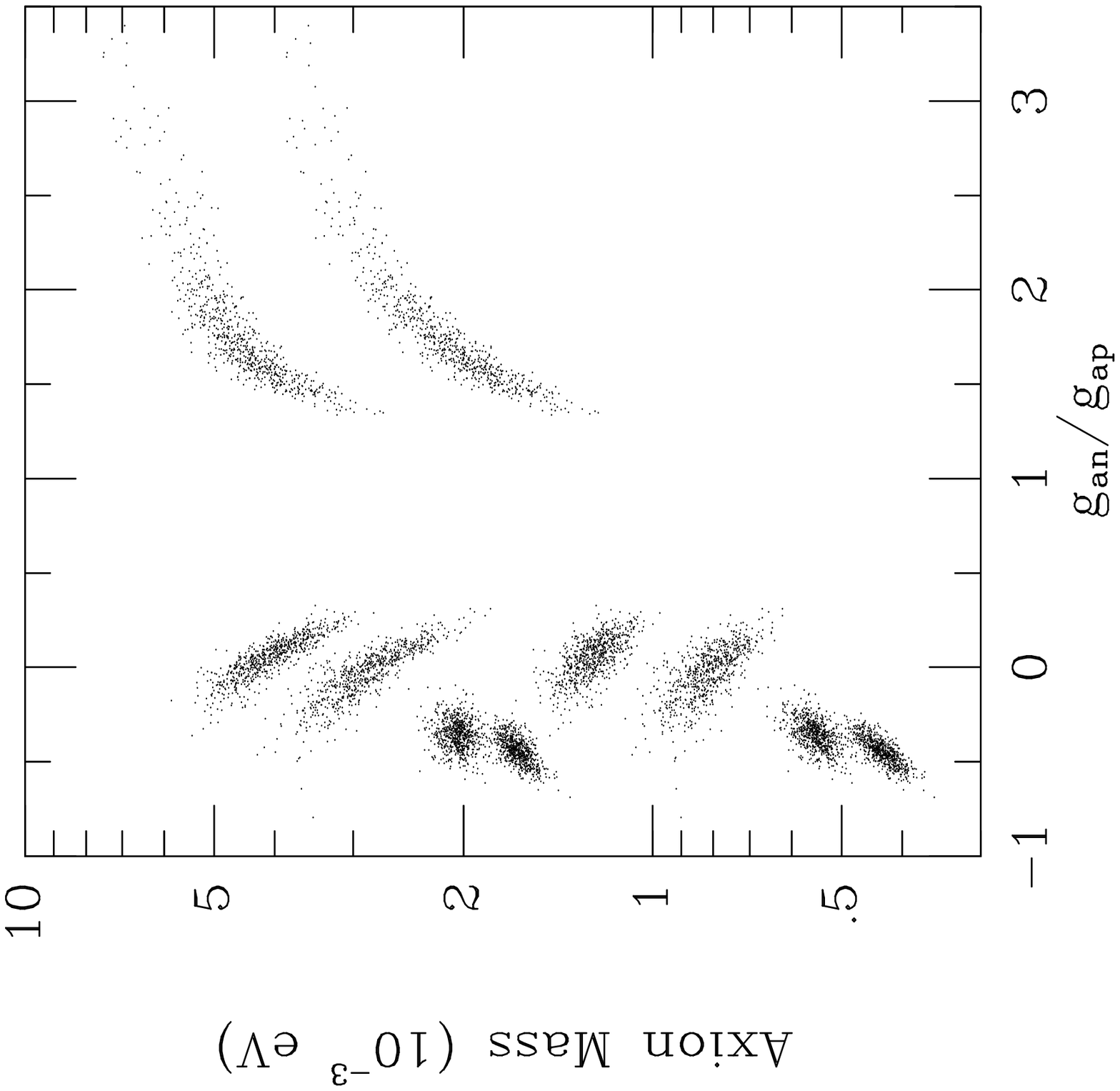}
\bigskip
\caption[...]{
SN~1987A limit to the axion mass as a
function of the ratio of axion-neutron to axion-proton couplings
for different axion models, with (lower set) and without
(upper set) pion processes; from bottom left to top right,
$\beta = 0$, $\beta = 27^\circ$, KSVZ, $\beta = 54^\circ$ and
$\beta = 81^\circ$. Note that saturation due to many-body
effects has not been included (it degrades the limit by a
factor of about 1.9).
\label{F13}}
\end{figure}

The upper bounds of the axion mass deduced for the points of 
Fig.~\ref{F1} by this procedure are plotted in Fig.~\ref{F13}. Above a line
connecting the point $(g_{an}/g_{ap}, m_a/[10^{-3}\,{\rm eV}]) = 
(-1,1)$ with the point $(3,5)$, Fig.~\ref{F13} shows the results for the 
``standard'' case without pions, whilst below this line the dots represent
the mass limits for the case with pion-axion conversion.
In both cases saturation effects have
been neglected. The latter case yields axion mass limits that
are stronger by a factor of almost four. As was mentioned
in Sect.~\ref{manyb}, the inclusion of saturation effects is expected
to lead to limits that are less sensitive to the presence of pions.

\section{Summary}

We have re-examined the stringent limit on the axion mass inferred
from neutrino emission by SN~1987A,
in the light of a possible suppression of axion emission
by the many-body effects of nucleon spin fluctuations and
additional emission processes involving pions,
taking into account the latest determinations
of axial-vector current couplings to nucleons.
The suppression of axion emission
due to many-body effects degrades previous limits by a factor of 
about 2. Emission processes involving thermal pions can strengthen
the limits by a factor of 3--4, if saturation effects on the
nucleon spin fluctuations are neglected, whereas inclusion
of such effects tends to make the limits less sensitive to pion
abundances. The resulting axion mass limit depends upon its precise
couplings, ranging from $0.5\times 10^{-3}\,{\rm eV}$
to $6\times 10^{-3}\,{\rm eV}$. Our results are consistent
with previous limits~\cite{burrowsetal}, though more
precisely stated.  Fig.~\ref{F12} shows our limit on the axion-proton
coupling as a function of the ratio of the axion-neutron
to axion-proton couplings, and Fig.~\ref{F13} shows the mass limit
as a function of this same ratio.

To test the stability of the deduced axion mass limits against
uncertainties of the stellar models and of the description of
the input physics, we performed a large number of comparative 
computations. It turned out that neither a change of the 
nucleon degeneracy parameter $\xi$ of Eq.~(\ref{thaverage})
nor a reduction of the neutrino opacity to 50\% of its standard
value lead to major changes of our conclusions. There are
differences of our work compared to previous work by 
Burrows et al.~\cite{burrowsetal}
concerning the protoneutron star modeling, namely we used 
different equations of state and did not take into account 
accretion of matter onto the nascent neutron star.

Two different equations of state were used in the presented
models, an $n$-$p$ equation of state and an equation of state where
hyperons occur as additional hadronic degrees of freedom at 
densities beyond about two times nuclear matter density \cite{glen}. 
For neutron stars with baryonic masses $\lesssim 1.70\,M_{\odot}$
hyperons are abundant only in a relatively small central core
region of the forming neutron star and therefore do not dramatically
change the cooling and neutrino emission of the star. Similarly,
the axion emission was found to be affected only at a minor level
by the presence of hyperons. 

The amount of material possibly accreted onto the protoneutron 
star during the first moments of the supernova explosion 
and the corresponding accretion rate as a function of 
time are rather uncertain and must depend on the structure of the
progenitor star and on the details of the
still incompletely understood explosion mechanism of Type II 
supernovae. 
Instead of introducing new free parameters to model accretion, we 
tried to account for the unknown amount of accreted matter by
performing cooling simulations for protoneutron star models
with baryonic masses between $1.3\,M_{\odot}$ and $1.75\,M_{\odot}$.
Interestingly, the higher temperatures in more massive stars lead
to enhanced axion emission that carries away a larger fraction
of the gravitational binding energy that is released during the
cooling of the star. Therefore the predicted neutrino signals 
in the KII and IMB detectors and thus the axion mass limits 
deduced from the SN~1987A neutrino data should be rather 
insensitive to the exact mass of the protoneutron star formed
in SN~1987A. Including accretion as in
Burrows et al.~\cite{burrowsetal} raises the total neutrino luminosity
above the contribution from the core and leads to additional
events in the detectors which are essentially unaffected by the 
emission of axions. Therefore the event numbers $N_{\rm KII}$ and
$N_{\rm IMB}$ in reference \cite{burrowsetal} were found to drop
somewhat less strongly with increasing axion-nucleon coupling
than they do in case of our models.

\section*{Acknowledgments.}

We thank Donald Q. Lamb and G.~Raffelt 
for many valuable discussions and comments. Useful remarks by M.~Ruffert
are also acknowledged. WK and HTJ are very grateful to N.K.~Glendenning for
providing them with the tables of his equations of state. They also want to
thank J.R.~Wilson for the data of the protoneutron star from one of his
supernova simulations, which was used to construct some of the initial
models for the investigations. This work was supported by
the DoE (at Chicago and Fermilab), by the NASA (at Fermilab
by grant NAG 5-2788), and by the ``Sonderforschungsbereich 375-95
f\"ur Astro-Teilchenphysik'' of the Deutsche Forschungsgemeinschaft
(at Garching). For a part of the work done during a stay at the
University of Chicago, HTJ acknowledges support 
by the NSF grant AST 92-17969, by the NASA
grant NAG 5-2081, and by an Otto Hahn Postdoctoral Scholarship
of the Max-Planck-Society. The computations were performed on the
CRAY-YMP 4/64 and Cray-EL98~4/256 of the Rechenzentrum Garching.

\end{document}